\documentclass[aip,rsi,superscriptaddress,fleqn,reprint]{revtex4-1}

\usepackage{amsmath, amsthm, amssymb}
\usepackage{graphicx}
\graphicspath{{figures/}}
\usepackage{epstopdf}
\usepackage{natbib}
\usepackage{dcolumn}

\begin{document}

\newcommand{\figtitle}{First line of caption, will appear in LOF}
\newcommand{\apex}{APEX-SZ}
\newcommand{\sz}{Sunyaev-Zel'dovich}
\newcommand{\veff}{\ensuremath{\Delta\nu_{\text{eff}}}}
\newcommand{\bst}{\ensuremath{3\lambda/4}}
\newcommand{\bso}{\ensuremath{\lambda/4}}
\newcommand{\het}{\ensuremath{^3\mbox{He}}}
\newcommand{\hef}{\ensuremath{^4\mbox{He}}}
\newcommand{\lcdm}{\ensuremath{\Lambda\text{CDM}}}
\newcommand{\sci}[1]{\ensuremath{\times 10^{#1}}}
\newcommand{\dcoltext}[1]{\multicolumn{1}{c}{\text{#1}}}
\newcommand{\rj}{\text{RJ}}

\newcommand{\oml}{\ensuremath{\Omega_\Lambda}}
\newcommand{\omm}{\ensuremath{\Omega_m}}
\newcommand{\omr}{\ensuremath{\Omega_r}}
\newcommand{\omk}{\ensuremath{\Omega_k}}
\newcommand{\sigeight}{\ensuremath{\sigma_8}}
\newcommand{\w}{\ensuremath{w}}

\newcommand{\ohm}{\ensuremath{\Omega}}
\newcommand{\rthz}{\ensuremath{\sqrt{\text{Hz}}}}
\newcommand{\uks}{\ensuremath{\mu\text{K}\sqrt{s}}}
\newcommand{\ukcmb}{\ensuremath{\mu\text{K}_\text{CMB}}}
\newcommand{\ukcmbs}{\ensuremath{\mu\text{K}_\text{CMB}\sqrt{s}}}
\newcommand{\amin}{\ensuremath{^\prime}}
\newcommand{\sqamin}{\ensuremath{\text{amin}^2}}
\newcommand{\asec}{\ensuremath{^{\prime\prime}}}
\newcommand{\dg}{\ensuremath{^\circ}}
\newcommand{\skypatch}[2]{\ensuremath{#1^\circ\times#2^\circ}}
\newcommand{\db}[1]{\ensuremath{#1\,\text{dB}}}
\newcommand{\mapspeed}{\ensuremath{\text{deg}^2\text{/mJy}^2\text{/hr}}}

\newcommand{\figsize}{height=3.5in}

\newcommand{\tab}{\hspace{5mm}}
\newcommand{\blankpage}{\clearpage ~ \newpage}

\newcommand{\abs}[1]{\left\vert#1\right\vert}
\newcommand{\set}[1]{\left\{#1\right\}}
\newcommand{\Real}{\mathbb R}
\newcommand{\eps}{\varepsilon}
\newcommand{\To}{\longrightarrow}

\newcommand{\oscdemod}{oscillator/demodulator}
\newcommand{\fMUX}{fMUX}
\newcommand{\squid}{SQUID}
\newcommand{\phinot}{\mbox{$\Phi_0$}}
\newcommand{\Likelihood}{\mbox{\L}}
\newcommand{\logLikelihood}{\mbox{$-\ln\Likelihood$}}
\newcommand{\fortran}{{\tt Fortran~77}}
\newcommand{\CXX}{C++}
\newcommand{\order}{\ensuremath{\cal O}}
\newcommand{\const}{\mbox{\sc\small Const}}
\newcommand{\mycomment}[1]{{\bf\it\color{red} #1}} 
\renewcommand{\arraystretch}{1.25}

\newcommand{\aj}{Astronomical Journal}
\newcommand{\araa}{Annual Review of Astron and Astrophys}
\newcommand{\apj}{Astrophysical Journal}
\newcommand{\apjl}{Astrophysical Journal, Letters}
\newcommand{\apjs}{Astrophysical Journal, Supplement}
\newcommand{\ao}{Applied Optics}
\newcommand{\apss}{Astrophysics and Space Science}
\newcommand{\aap}{Astronomy and Astrophysics}
\newcommand{\aapr}{Astronomy and Astrophysics Reviews}
\newcommand{\aaps}{Astronomy and Astrophysics, Supplement}
\newcommand{\azh}{Astronomicheskii Zhurnal}
\newcommand{\baas}{Bulletin of the AAS}
\newcommand{\jrasc}{Journal of the RAS of Canada}
\newcommand{\memras}{Memoirs of the RAS}
\newcommand{\mnras}{Monthly Notices of the RAS}
\newcommand{\pra}{Physical Review A: General Physics}
\newcommand{\prb}{Physical Review B: Solid State}
\newcommand{\prc}{Physical Review C}
\newcommand{\prd}{Physical Review D}
\newcommand{\pre}{Physical Review E}
\newcommand{\prl}{Physical Review Letters}
\newcommand{\pasp}{Publications of the ASP}
\newcommand{\pasj}{Publications of the ASJ}
\newcommand{\qjras}{Quarterly Journal of the RAS}
\newcommand{\skytel}{Sky and Telescope}
\newcommand{\solphys}{Solar Physics}
\newcommand{\sovast}{Soviet Astronomy}
\newcommand{\ssr}{Space Science Reviews}
\newcommand{\zap}{Zeitschrift fuer Astrophysik}
\newcommand{\nat}{Nature}
\newcommand{\iaucirc}{IAU Cirulars}
\newcommand{\aplett}{Astrophysics Letters}
\newcommand{\apspr}{Astrophysics Space Physics Research}
\newcommand{\bain}{Bulletin Astronomical Institute of the Netherlands}
\newcommand{\fcp}{Fundamental Cosmic Physics}
\newcommand{\gca}{Geochimica Cosmochimica Acta}
\newcommand{\grl}{Geophysics Research Letters}
\newcommand{\jcp}{Journal of Chemical Physics}
\newcommand{\jgr}{Journal of Geophysics Research}
\newcommand{\jqsrt}{Journal of Quantitiative Spectroscopy and Radiative Transfer}
\newcommand{\memsai}{Mem. Societa Astronomica Italiana}
\newcommand{\nphysa}{Nuclear Physics A}
\newcommand{\physrep}{Physics Reports}
\newcommand{\physscr}{Physica Scripta}
\newcommand{\planss}{Planetary Space Science}
\newcommand{\procspie}{Proceedings of the SPIE}

\title{Millimeter-wave bolometer array receiver for the Atacama pathfinder experiment Sunyaev-Zel'dovich (\apex) instrument}

\author{D.~Schwan}
\email{schwan@berkeley.edu}
\affiliation{Department of Physics, University of California, Berkeley, CA, 94720}
\author{P.~A.~R.~Ade}
\affiliation{School of Physics and Astronomy, Cardiff University, CF24 3YB Wales, UK}
\author{K.~Basu}
\affiliation{Argelander Institute for Astronomy, Bonn University, Bonn, Germany}
\author{A.~N.~Bender}
\affiliation{Center for Astrophysics and Space Astronomy, Department of Astophysical and
Planetary Sciences, University of Colorado, Boulder, CO, 80309}
\author{F.~Bertoldi}
\affiliation{Argelander Institute for Astronomy, Bonn University, Bonn, Germany}
\author{H.-M.~Cho}
\affiliation{National Institute of Standards and Technology, Boulder, CO, 80305}
\author{G.~Chon}
\affiliation{Max Planck Institute for Extraterrestrial Physics, 85748 Garching, Germany}
\author{John~Clarke}
\affiliation{Department of Physics, University of California, Berkeley, CA, 94720}
\affiliation{Materials Sciences Division, Lawrence Berkeley National Laboratory, Berkeley, CA, 94720}
\author{M.~Dobbs}
\affiliation{Department of Physics, McGill University, Montr\'{e}al, Canada, H3A 2T8}
\author{D.~Ferrusca}
\affiliation{Department of Physics, University of California, Berkeley, CA, 94720}
\author{R.~G\"usten}
\affiliation{Max Planck Institute for Radio Astronomy, 53121 Bonn, Germany}
\author{N.~W.~Halverson}
\affiliation{Center for Astrophysics and Space Astronomy, Department of Astophysical and
Planetary Sciences, University of Colorado, Boulder, CO, 80309}
\affiliation{Department of Physics, University of Colorado, Boulder, CO, 80309}
\author{W.~L.~Holzapfel}
\affiliation{Department of Physics, University of California, Berkeley, CA, 94720}
\author{C.~Horellou}
\affiliation{Onsala Space Observatory, Chalmers University of Technology, SE-439 92 Onsala, Sweden}
\author{D.~Johansson}
\affiliation{Onsala Space Observatory, Chalmers University of Technology, SE-439 92 Onsala, Sweden}
\author{B.~R.~Johnson}
\affiliation{Department of Physics, University of California, Berkeley, CA, 94720}
\author{J.~Kennedy}
\affiliation{Department of Physics, McGill University, Montr\'{e}al, Canada, H3A 2T8}
\author{Z.~Kermish}
\affiliation{Department of Physics, University of California, Berkeley, CA, 94720}
\author{R.~Kneissl}
\affiliation{European Southern Observatory, Alonso de C\'{o}rdova 3107, Vitacura, Santiago, Chile}
\affiliation{Atacama Large Millimeter Array Joint ALMA Observatory, Av. El Golf 40 - Piso 18, Las Condes, Santiago, Chile}
\author{T.~Lanting}
\affiliation{Department of Physics, McGill University, Montr\'{e}al, Canada, H3A 2T8}
\author{A.~T.~Lee}
\affiliation{Department of Physics, University of California, Berkeley, CA, 94720}
\affiliation{Physics Division, Lawrence Berkeley National Laboratory, Berkeley, CA, 94720}
\author{M.~Lueker}
\affiliation{Department of Physics, University of California, Berkeley, CA, 94720}
\author{J.~Mehl}
\affiliation{Department of Physics, University of California, Berkeley, CA, 94720}
\author{K.~M.~Menten}
\affiliation{Max Planck Institute for Radio Astronomy, 53121 Bonn, Germany}
\author{D.~Muders}
\affiliation{Max Planck Institute for Radio Astronomy, 53121 Bonn, Germany}
\author{F.~Pacaud}
\affiliation{Argelander Institute for Astronomy, Bonn University, Bonn, Germany}
\author{T.~Plagge}
\affiliation{Department of Physics, University of California, Berkeley, CA, 94720}
\author{C.~L.~Reichardt}
\affiliation{Department of Physics, University of California, Berkeley, CA, 94720}
\author{P.~L.~Richards}
\affiliation{Department of Physics, University of California, Berkeley, CA, 94720}
\author{R.~Schaaf}
\affiliation{Argelander Institute for Astronomy, Bonn University, Bonn, Germany}
\author{P.~Schilke}
\affiliation{Max Planck Institute for Radio Astronomy, 53121 Bonn, Germany}
\author{M.~W.~Sommer}
\affiliation{Argelander Institute for Astronomy, Bonn University, Bonn, Germany}
\affiliation{Max Planck Institute for Radio Astronomy, 53121 Bonn, Germany}
\author{H.~Spieler}
\affiliation{Physics Division, Lawrence Berkeley National Laboratory, Berkeley, CA, 94720}
\author{C.~Tucker}
\affiliation{School of Physics and Astronomy, Cardiff University, CF24 3YB Wales, UK}
\author{A.~Weiss}
\affiliation{Max Planck Institute for Radio Astronomy, 53121 Bonn, Germany}
\author{B.~Westbrook}
\affiliation{Department of Physics, University of California, Berkeley, CA, 94720}
\author{O.~Zahn}
\affiliation{Department of Physics, University of California, Berkeley, CA, 94720}

\begin{abstract}
The Atacama pathfinder experiment \sz\ (\apex) instrument is a millimeter-wave cryogenic receiver designed to observe galaxy clusters via the \sz\ effect from the 12~m APEX telescope on the Atacama plateau in Chile.  The receiver contains a focal plane of 280 superconducting transition-edge sensor (TES) bolometers instrumented with a frequency-domain multiplexed readout system.  The bolometers are cooled to 280~mK via a three-stage helium sorption refrigerator and a mechanical pulse-tube cooler.  Three warm mirrors, two 4~K lenses, and a horn array couple the TES bolometers to the telescope.  \apex\ observes in a single frequency band at 150~GHz with 1\amin\ angular resolution and a 22\amin\ field-of-view, all well suited for cluster mapping.  The \apex\ receiver has played a key role in the introduction of several new technologies including TES bolometers, the frequency-domain multiplexed readout, and the use of a pulse-tube cooler with bolometers.  As a result of these new technologies, the instrument has a higher instantaneous sensitivity and covers a larger field-of-view than earlier generations of \sz\ instruments.  The TES bolometers have a median sensitivity of 890~\ukcmbs\  (NE$y$ of 3.5\sci{-4}~$\sqrt{s}$).  We have also demonstrated upgraded detectors with improved sensitivity of 530~\ukcmbs\ (NE$y$ of 2.2\sci{-4}~$\sqrt{s}$).  Since its commissioning in April 2007, \apex\ has been used to map 48 clusters.   We describe the design of the receiver and its performance when installed on the APEX telescope.
\\\vspace{0.0625 in}\\
\textit{Copyright 2011 American Institute of Physics. This article may be downloaded for personal use only. Any other use requires prior permission of the author and the American Institute of Physics.}\\
\vspace{0.0625 in}\\\textit{The following article appeared in} D. Schwan et al., Rev. Sci. Inst., \textbf{82}, 091301 (2011) \textit{and may be found at}  \url{http://link.aip.org/link/?rsi/82/091301}.
\end{abstract}


\maketitle


\section{Introduction}

Galaxy clusters are the most massive gravitationally collapsed objects in the Universe, offering a unique probe into the composition and evolution of the universe (see, e.g., Refs.~\onlinecite{barbosa1996,bahcall1998}).  Clusters of galaxies formed as dark energy became the dominant component of  energy density in the universe.  The growth of structure from this period on is sensitive to the dark energy density and its equation of state.  Therefore, measuring the evolution of cluster number density is a powerful method to constrain these parameters as well as the matter power spectrum normalization and the mass density of the universe.\cite{holder2001}  In addition, galaxy clusters are important laboratories for studying various astrophysical phenomena in great detail, like heat transport and instabilities in magnetized plasmas,\cite{parrish2008} evolution of galaxies in dense environments,\cite{delucia2007} and cosmic circulation of heavy elements.\cite{kapferer2007}

The bulk of the baryonic mass in galaxy clusters resides in the form of a hot, low-density ionized gas filling the space between its constituent galaxies.  This intracluster plasma is observable at mm-wavelengths via the \sz\ effect (SZE),\cite{sunyaev1970,*sunyaev1972} a distortion of the cosmic microwave background (CMB) blackbody spectrum arising from inverse Compton scattering of $\sim1\%$ of the CMB photons by intracluster electrons.  At 150~GHz, the SZE blackbody distortion is visible as an intensity decrement in the CMB.
The surface brightness of the SZE signal is proportional to the integrated gas pressure through the cluster. The signal is independent of cluster redshift, since the SZE produces a fractional change in the CMB intensity which remains constant as the light is redshifted between cluster and observer. In contrast, x-ray and optical surface brightness dim with increasing redshift.  The redshift independence makes the SZE a uniquely sensitive method for discovering and observing distant clusters (see, e.g., Refs.~\onlinecite{birkinshaw1999,carlstrom2002}).

The Atacama pathfinder experiment \sz\ (\apex) instrument is a millimeter-wave receiver for the APEX telescope designed to make sensitive measurements of the SZE in galaxy clusters.\cite{schwan2003}  The \apex\ focal plane contains a 280 element superconducting transition-edge sensor (TES) bolometer array operating at 280~mK.  The receiver-telescope combination has 1\amin\ resolution at its observation frequency of 150~GHz and a 22\amin\ field-of-view.  Specifications and measured performance for \apex\ are summarized in Table~\ref{tbl:apexspec}.

\begin{table}[t]\centering
\begin{ruledtabular}
\begin{tabular}{lcc}
  \em{Optics} & \\
  Primary mirror physical diameter (m) & \multicolumn{2}{c}{12}\\
  Primary mirror illumination diameter (m) & \multicolumn{2}{c}{8}\\
  Field-of-view (amin) & \multicolumn{2}{c}{22}\\
  Beam FWHM (asec) & \multicolumn{2}{c}{58}\\
 \\
  \em{Detectors} & \\
  Bolometers & \multicolumn{2}{c}{280}\\
  Live channels & \multicolumn{2}{c}{170--180} \\
  \\
  \em{Performance} & \em{Type-1} & \em{Type-2}\\
  Band center (GHz) & 151 & 154\\
  Bandwidth ($\Delta \nu$) (GHz) & 24.5 & 33.5 \\
  Cumulative efficiency ($\eta_{tot}$) & 0.31 & 0.36\\
  NET (\ukcmbs)	&	890 & 530\\
  NE$y$ (\sci{-4}~$\sqrt{s}$) & 3.5 & 2.2\\
\end{tabular}
\end{ruledtabular}
\caption{\apex\ specifications and measured performance.  The number of detectors typically used in data analysis are listed as ``live channels''  (see Sec.~\ref{sec:perf.detect}).  Bolometer performance values are listed for both type-1 and type-2 detectors (Sec.~\ref{sec:perf}).  The NET and NE$y$ are listed for a single detector and refers to the median performance of the array (Sec.~\ref{sec:perf.sens}).}
\label{tbl:apexspec}
\end{table}

The \apex\ detector array comprises six triangular sub-arrays with 55 bolometers each.  The receiver was first installed on APEX with a single sub-array in December 2005 for instrument characterization.  We redeployed the receiver with a full focal plane array of 280 multiplexed detectors in April 2007. Since then, we have accrued 875~h of observations in seven observing runs, using observing time allotted to German and Swedish institutions.  Over this period, \apex\ has used two types of detectors with different absorption efficiencies and thermal links.  We refer to these as type-1 and type-2; Sec.~\ref{sec:optics.focal} details the differences between the two detector designs.

\apex\ focuses on wide-field imaging of individual clusters to probe the relationship between SZE flux and cluster mass at varying redshift.  It is one of a number of millimeter-wave bolometer instruments which have come online recently.  The Large APEX Bolometer Camera (LABOCA; Ref.~\onlinecite{siringo2009}) and AzTEC\cite{wilson2008b} both observe from the Atacama at 345 and 270~GHz, respectively.  Observations of clusters at 150~GHz and higher frequencies are complementary, providing additional insight to cluster dynamics and  constraining contamination from other sources.  The South Pole Telescope (SPT; Ref.~\onlinecite{ruhl2004,*carlstrom2011} and the Atacama Cosmology Telescope (ACT; Ref.~\onlinecite{kosowsky2003,*swetz2011}) are two dedicated SZE survey instruments which have discovered hundreds of clusters.\cite{vanderlinde2010,*menanteau2010} \apex\ is more suited for targeted observation of clusters for multi-frequency study.  The Atacama location enables SZE study of existing x-ray data fields which is limited from the South Pole.   The ACT is designed for surveying and does not have an elevation drive for pointing at individual clusters.  The SZE-mass relationship measured by \apex\ will be critical for developing the cosmological constraints from the results of these surveys.

SZE observations are functionally CMB anisotropy measurements at high sensitivity and resolution, and \apex\ represents a generational leap from the bolometric receivers that measured the primary anisotropies at degree scales.  These receivers (such as the Millimeter wave Anisotropy eXperiment IMaging Array (MAXIMA; Ref.~\onlinecite{rabii2006}) and Balloon Observations Of Millimetric Extragalactic Radiation and Geophysics (BOOMERanG; Ref.~\cite{crill2003}) each with 16 detectors) used neutron-transmutation-doped (NTD) bolometers.  The bolometers were housed in individual integrating cavities with separate feed horns and filters.  Developed later, the Bolocam\cite{glenn1998} receiver employs a focal plane array of 144 NTD bolometers, in which the absorbers are all built lithographically on a single silicon wafer.  The focal plane includes a feed horn array and a single, large filter.

In constructing \apex, we have expended much effort in developing robust, scalable technologies---notably TES bolometers and frequency-domain multiplexed SQUID readout compatible with the pulse tube cooler---which represent a significant advance and enable a large increase in sensitivity compared with previous bolometric receivers.   The bolometers are fabricated on silicon wafers with thin-film deposition and optical lithography.  The focal plane and readout are assembled from several modular components enabling scaling to larger arrays.   These components have been used in SPT, which started observations in 2007 with 960 detectors, and \textsc{Polarbear},\cite{arnold2010} an upcoming Berkeley CMB polarization experiment with over 1200 detectors. The analog multiplexed SQUID readout developed for \apex\ has evolved into a next-generation digital multiplexed readout system described in Ref.~\onlinecite{dobbs2008}.  The digital multiplexer will be used in \textsc{Polarbear}, SPTpol,\cite{mcmahon2009} and the E~and~B Experiment (EBEX; Ref.~\onlinecite{aubin2010}).

This paper describes the design and performance of \apex.  The structure of the paper is as follows: the APEX telescope and site are discussed in Sec.~\ref{sec:apex}, optics in Sec.~\ref{sec:optics}, TES detectors in Sec.~\ref{sec:yield}, multiplexed readout in Sec.~\ref{sec:readout}, cryostat and cooling systems in Sec.~\ref{sec:cryogenics}, system performance in Sec.~\ref{sec:perf}, observations in Sec.~\ref{sec:obs}, and summary and conclusions in Sec.~\ref{sec:conc}.

\section{APEX}
\label{sec:apex}

\subsection{Telescope}
\begin{figure}[th]\centering
\includegraphics{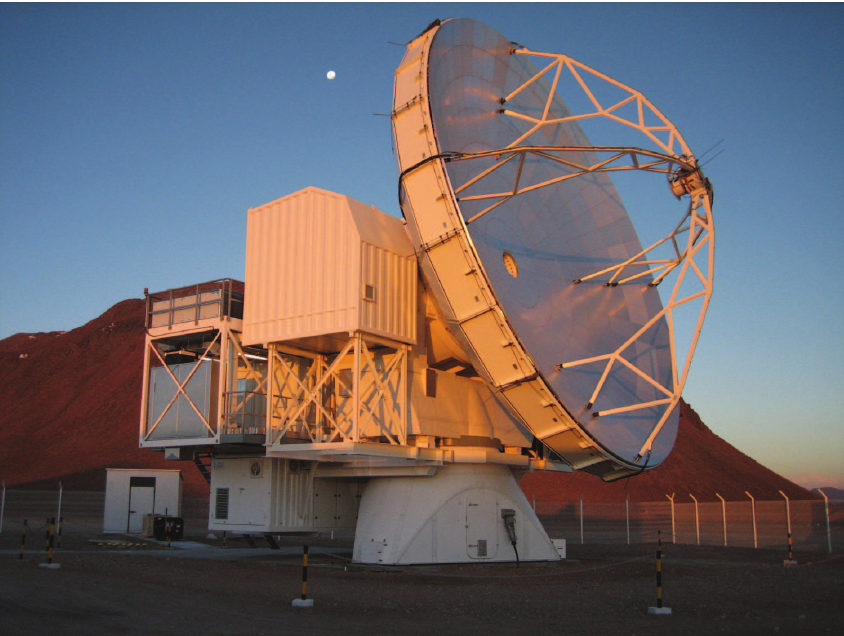}
    \caption[]{The 12~m APEX telescope on Llano de Chajnantor at 5107~m altitude.}
    \label{fig:apex}
\end{figure}

The APEX telescope (shown in Fig.~\ref{fig:apex}) is a 12-meter diameter on-axis Cassegrain telescope based on the design of the VERTEX Atacama Large Millimeter Array (ALMA) prototype, but modified to include two Nasmyth cabins.\cite{gusten2006} the APEX was commissioned by the Max Planck Institut f\"{u}r Radioastronomie, the European Southern Observatory, and the Swedish Onsala Space Observatory for use with bolometric and heterodyne receivers and began operations in September 2005.

The primary mirror consists of 264 panels aligned via radio holography to form a surface with 18~$\mu$m rms accuracy allowing observations up to 1.5~THz.  The secondary is 0.75~m in diameter and is mounted on a hexapod for optical adjustment.  It focuses through a 0.75~m hole in the primary, which limits the field-of-view (FOV) to 22\amin.   This FOV allows \apex\ to map the large angular size of low-redshift clusters.  The primary and secondary mirrors have been etched to scatter infrared and optical light, enabling daylight observations.  The absolute pointing accuracy is measured to be 2\asec\ rms.  The APEX secondary support legs are made from carbon-fiber reinforced plastic and tapered to minimize scattering, and the large size of the primary mirror allows it to be under-illuminated while still achieving arcminute resolution.  These two features are important for minimizing ground contamination since the telescope lacks a ground shield.

\subsection{Atacama Site}
\label{sec:site}
The APEX site is at 5107~m altitude on Llano de Chajnantor in the Atacama Desert in northern Chile.  The Atacama is one of the driest areas on the Earth and considered, along with the South Pole and Mauna Kea, to be one of the premier locations for millimeter and sub-millimeter wave astronomy.

The APEX is located near the ALMA site and within a few kilometers of the sites of a number of past (TOCO; Ref.~\onlinecite{miller2002} and CBI; Ref.~\onlinecite{padin2002}) and current (ACT; Ref.~\onlinecite{kosowsky2003} and QUIET; Ref.~\onlinecite{buder2010}) CMB experiments.  The ALMA site testing campaign made atmospheric measurements at 225~GHz over several years and found atmospheric opacity at Chajnantor and the South Pole to be comparable with median zenith optical depths of $\tau_{225}=0.061$ and $\tau_{225}=0.053$, respectively.\cite{chamberlin1994,*chamberlin1995,*radford2000}   While the opacities are comparable, the relative contribution from water vapor is greater in the Atacama desert than at the South Pole.
Spatial variation of emission from water vapor results in  higher fluctuation power at Chajnantor.\cite{lay2000,*bussmann2005}
Diurnal and annual variations in weather at Chajnantor can have a
significant impact on the observed atmospheric noise.  For much of the year, conditions are excellent, but during much of the Austral summer, the ``Bolivian winter" weather phenomenon brings 2--3 months of increased atmospheric opacity and frequent precipitation.\cite{otarola2005}

In addition to favorable observing conditions, the low latitude of the APEX site (23:00:20.8S, 67:45:33.0W) allows observations of a large fraction of the celestial sphere.  A large number of fields with rich multi-frequency data sets not accessible from the South Pole are observable from Chajnantor.
Overlapping observations at several wavelengths enable a more complete study and characterization of clusters.  For example, cluster observations within the x-ray Multi-Mirror Mission-Large Scale Survey (XMM-LSS) field offer immediate comparison with x-ray data, and the Cosmological Evolution Survey (COSMOS) field provides a rich set of weak lensing, x-ray, and sub-millimeter data to cross-calibrate and characterize foregrounds for SZE observations.

\section{Optics}
\label{sec:optics}

\subsection{Tertiary Optics}
\label{sec:tertoptics}

The \apex\ tertiary optics couple the telescope beam to the focal plane, which is an array of feed-horn coupled bolometers.  The optical system has diffraction limited performance over the full 22\amin\ FOV of APEX, and the focal plane is flat and telecentric, with $f/2.3$.  An additional constraint on the optical system is that the pulse-tube cooler cooling power varies with orientation (Sec.~\ref{sec:ptc}). The cryostat optical axis is tilted 30\dg\ from the telescope optical axis to keep the pulse-tube cooler within 30\dg\ of vertical over a 60\dg\ range in elevation angle.

\begin{figure}[th]\centering
\includegraphics{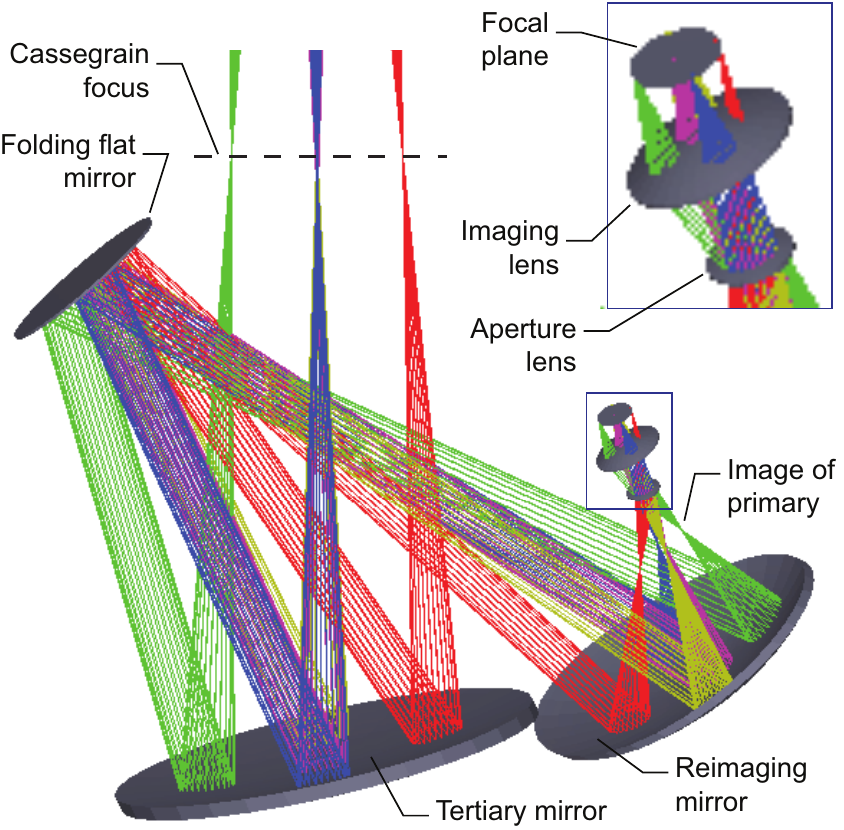}
    \caption[]{The \apex\ reimaging optics design. Each bundle of rays represents the beam from a single detector.}
    \label{fig:optics}
\end{figure}

Figure~\ref{fig:optics} illustrates the \apex\ tertiary optical system, which comprises three ambient temperature mirrors, a 4~K Lyot stop, and two 4~K high-density polyethylene (HDPE) lenses.  The tertiary mirror is a 1440~mm diameter off-axis paraboloid on the cabin floor.  The tertiary is followed by a 620~mm flat (quaternary) and a 980~mm $\times$ 1160~mm ellipsoidal mirror (quinary) that creates an image of the primary inside the cryostat. All three mirrors are pure conic sections.  The cryostat window aperture is 150~mm in diameter and is made from a 50~mm thick layer of Zotefoam PPA30\footnote{Zotefoams plc, 675 Mitcham Road, Croydon CR9 3AL, UK} epoxied into an aluminum ring. Zotefoam is a nitrogen-expanded polypropylene foam, which has low scattering and absorption at millimeter wavelengths.  We measure the transmittance to be greater than 99\% averaged across the instrument frequency response band.

The 72~mm diameter 4~K Lyot stop is located at the primary image, and it truncates the beam at $\sim 0.7$ of the primary diameter to prevent spillover and ground contamination without significantly increasing the bolometer optical load. The edge taper at the Lyot stop is $-4.5$~dB, corresponding to a fractional spillover power of $0.34$ for the $D = 1.4 f\lambda$ diameter conical horns, where $f$ is the focal ratio and $\lambda$ is the optical wavelength. The 1\amin\ FWHM beams on the sky are the far-field diffraction pattern of this illumination profile. The horn diameter, and correspondingly, the Lyot stop edge taper and spillover power, were chosen to maximize array mapping speed given two design constraints: a maximum 22\amin\ FOV and a maximum of 280 readout channels. (An unlimited FOV favors larger horns, whereas an unlimited number of readout channels favors smaller horns.) The Lyot stop is blackened with a 7~mm thick layer of Eccosorb MF-117,\footnote{Emerson and Cumming, 28 York Ave., Randolph, MA 02368} which is machined with triangular grooves to reduce reflections.

All optical elements on the sky-side of the Lyot stop are sized wherever possible to keep fractional spillover power below $0.001$ for a pixel on the edge of the array, see Table~\ref{tbl:optics}. This design minimizes both optical loading and truncation of diffractive sidelobes generated by the Lyot stop.  Such truncation would degrade the quality of the primary aperture stop. To assess the effect of finite optical element size on the beams and primary aperture stop, we modeled the beam propagation using multi-moded Gaussian beam propagation formalism and the \textsc{zemax} physical optics propagation software.\footnote{\textsc{zemax} Development Corp., 3001 112th Avenue NE, Suite 202, Bellevue, WA 98004}  Both methods gave similar results, and indicated that our optical element sizing was adequate.

Two 4~K high-density polyethylene (HDPE) lenses after the Lyot provide the final beam shaping. To maximize transmittance and minimize scattered light inside the cryostat, the HDPE lenses have machined antireflection surfaces with closely spaced, circumferential grooves which have triangular cross-sections 0.56~mm wide and 0.64~mm deep.\cite{halverson2002t}  In effect, these grooves produce a graded index surface coating between the vacuum and the HDPE, which has an index of refraction $n = 1.567$ at 4~K.\cite{lamb1996,*birch1984,*birch1993}  We calculate that the reflection loss at each lens surface should be less than~2\% averaged across the band.  The imaging lens axis is tilted with respect to the optical axis to compensate for the image tilt induced by the off-axis mirrors and create a flat (telecentric) focal plane.

\begin{table*}[th]\centering
\begin{ruledtabular}
\begin{tabular}{lccccc}
Optical Element & Diameter & Shape & Effective Focal & Gaussian Beam & Fractional \\
 & (mm) & & Length (mm) & Width (mm) & Spillover\\
\colrule
Teriary mirror & 1440 & Off-axis paraboloid & 1970 & 79 & 0.001 \\
Folding flat mirror & 620 & Flat & $\infty$ & 66 & 0.001 \\
Reimaging mirror & $920 \times 980$ & Ellipsoid & 559 & 58 & 0.003 \\
4~K Lyot stop & 72 & Circular aperture & $\infty$ & 45 & 0.34 \\
Aperture lens & 120 & Double convex & 213  & 47 & 0.07 \\
Imaging lens & 240 & Double convex & 175 & 36 & 0.01 \\
\end{tabular}
\end{ruledtabular}
\caption{\apex\ tertiary optics parameters. The optical elements are described in the text and in Fig.~\ref{fig:optics}. The Gaussian beam width is the radius at which the power in the fundamental Gaussian mode for a single pixel's beam drops to $e^{-2}$ of its peak power. The fractional spillover is fractional power vignetted by the optical element for a pixel on the edge of the focal plane array.}
\label{tbl:optics}
\end{table*}

\subsection{Filters and Band}
\label{sec:filters}

\begin{figure}[th]\centering
\includegraphics[width=3.375in]{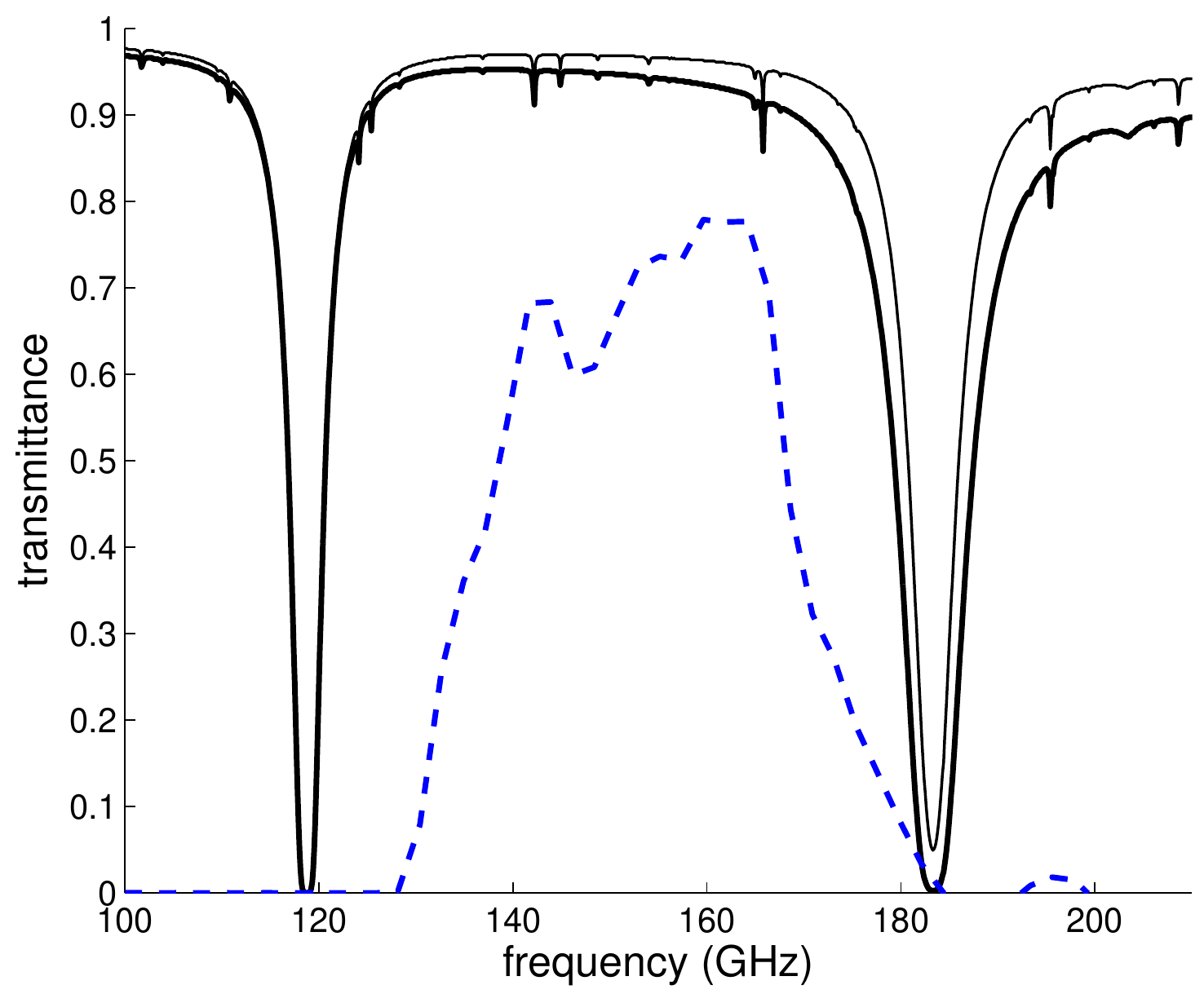}
    \caption[]{Calculated atmospheric transmittance at APEX for 2~mm (solid line) and 1~mm (thin solid line)  PWV (precipitable water vapor) at 55\dg\ elevation plotted with the measured \apex\ observation band (for type-2 detectors) (dashed line).  The atmospheric transmittance data are from the APEX transmittance calculator based on the ATM model.\cite{pardo2001}}
    \label{fig:band}
\end{figure}

\apex\ observes in a single band centered at 150~GHz near the maximum intensity decrement of the SZE at $\sim130$~GHz.  The measured observation pass-band (shown in Fig.~\ref{fig:band}) is selected to minimize contamination from other astronomical signals and avoid the atmospheric molecular absorption lines for oxygen at 118~GHz and water vapor at 183~GHz.  The pass-band is determined by a series of low-pass (LP) optical filters\cite{lee1996a} (described below), cylindrical waveguides and the bolometer absorption cavity (described in Sec.~\ref{sec:optics.focal}).  The broad water-vapor line at 183~GHz contributes roughly 3\% to the total background load on the bolometers (see Sec.~\ref{sec:perf.loading}).

\begin{figure}[th]\centering
\includegraphics{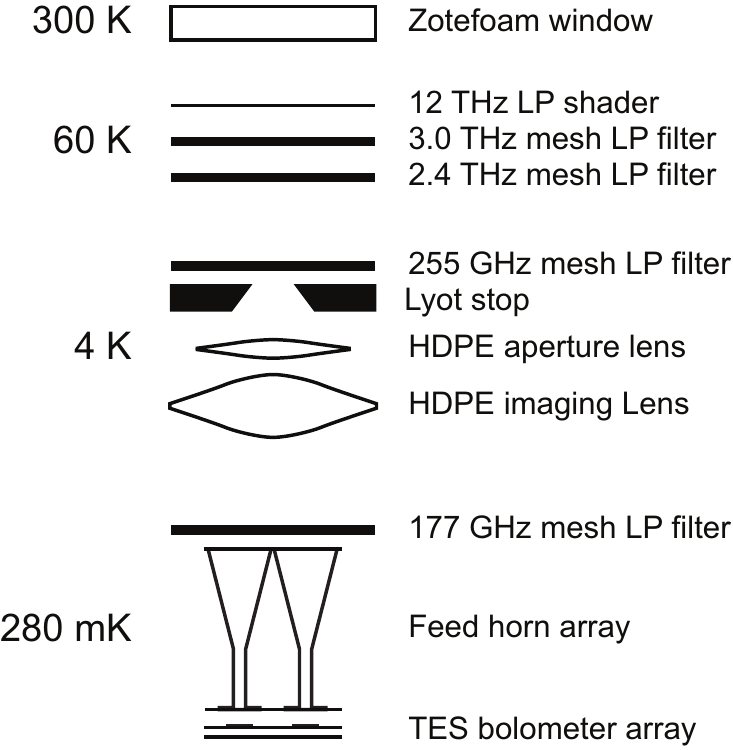}
    \caption[]{Schematic diagram showing optical components inside the \apex\ cryostat. Photons enter the cryostat via the Zotefoam window and pass through low-pass filters at 60~K and 4~K, two HDPE lenses, and a band-defining filter at 280~mK.  Finally, smooth-walled conical horns couple photons to the bolometers.}
    \label{fig:filterscheme}
\end{figure}

A series of LP optical filters serves to minimize radiative loading on the interior of the cryostat as well as define the observation band.  The LP filters consist of layers of capacitive, resonant metal-mesh in polyethylene.  The filter scheme is shown in Fig.~\ref{fig:filterscheme}.  The first element is a thin single layer LP filter, with metal-mesh on both sides of a polyethylene membrane, called an infrared shader.  It is designed for a slow cut-off with minimal scattering loss.  The $\sim$12~THz (400~$\mbox{cm}^{-1}$) LP shader reduces the radiative load on subsequent filters.     We minimize loading from 300~K emission on the 4~K stage with 3.0~THz (100~$\mbox{cm}^{-1}$) and 2.4~THz (80~$\mbox{cm}^{-1}$) multilayer LP filters at 60~K. Similarly, a 255~GHz (8.5~$\mbox{cm}^{-1}$) multilayer LP filter at the front of the 4~K radiation shell minimizes radiative loading from 60~K on the millikelvin stage.  The series of filters with staggered cutoffs also serves to block harmonic leaks that occur in metal-mesh filters.

The high-frequency edge of the observation band is defined by a 177~GHz (5.9~$\mbox{cm}^{-1}$) multilayer LP filter mounted directly on the feed horn array at 280~mK. The low-frequency edge of the observation band is defined by 1.33~mm diameter, 7.1~mm long cylindrical waveguides coupled to the feed horns described in Sec.~\ref{sec:optics.focal}.  Optical transmittance within the cutoffs defined by the filter and waveguide is further affected by the geometry of the bolometer absorption cavity.  Measurements of the achieved band with a Fourier transform spectrometer are described in Sec.~\ref{sec:perf.band}.

Each filter is clamped around its edge by an aluminum ring sandwiched with a beryllium-copper spring washer to compensate for differential thermal contraction of the aluminum and the polypropylene filter.  The washer maintains thermal contact between the filter and ring at low temperatures.  There is a 6--7~K temperature difference between the front of the 4~K radiation shell and the center of the 255~GHz filter. Though the temperature gradient is large, the filter does not add significant thermal load to the detectors, and the gradient itself is not a concern.

\subsection{Focal Plane Optics}
\label{sec:optics.focal}

\begin{figure}[th]\centering
\includegraphics{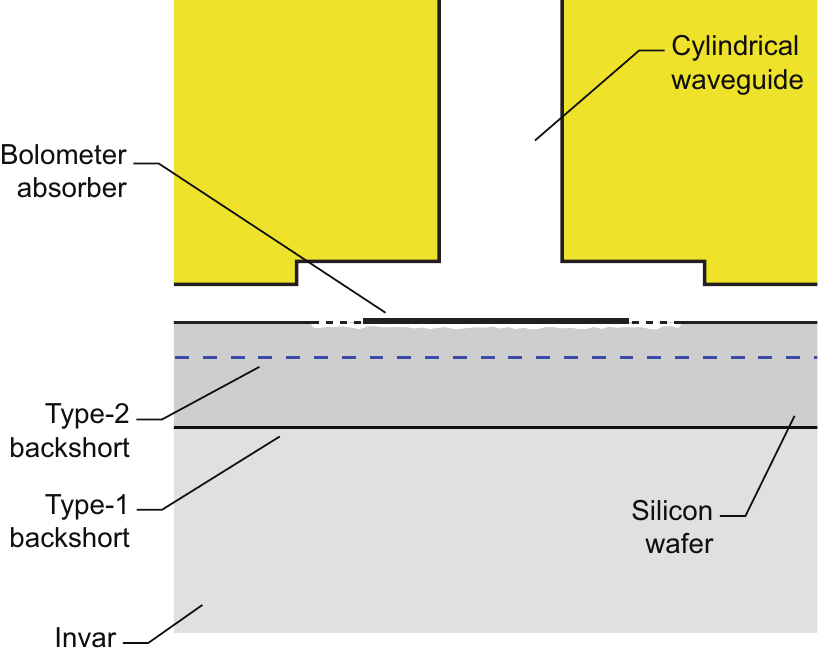}
    \caption[]{Schematic diagram of the \apex\ integrating cavity geometry for a single bolometer.  The bolometers are lithographed on a silicon wafer and centered behind an array of cylindrical waveguides that couple to the smooth-wall horns.  The horn/waveguide array is positioned $400~\mu$m from the wafer, and there is a 4~mm diameter, $250~\mu$m deep relief at the end of the waveguide which reduces radiation loss along the wafer.  The bolometer absorber is suspended on a silicon nitride membrane above a $20~\mu$m vacuum gap etched in the silicon.  The wafer is mounted on invar.  In type-1 detectors, the invar mount serves as a \bst\ backshort.  In type-2 detectors, two thin wafers are joined with a metal film between them to serve as a \bso\ backshort.  The wafer and bolometer thicknesses are exaggerated for clarity.}
    \label{fig:cavity}
\end{figure}

The \apex\ focal plane consists of 330 feed-horn coupled bolometers. Of these, 280 detectors are read out (see Sec.~\ref{sec:readout}).  The horn array mounted above the bolometer wafer is machined from a single aluminum block then gold plated.  The horns are arranged in a hexagonal close-packed configuration with a 6.7~mm ($1.4 f\lambda$) spacing so that the entire focal plane is 133~mm in diameter.  Figure~\ref{fig:cavity} shows a schematic diagram of the \apex\ integrating cavity; the design is similar to that of the Bolocam array.\cite{glenn2002}

The detectors are thermally isolated, suspended structures built upon a silicon wafer which is mounted on an invar plate.  \apex\ has used two types of detectors which differ in the location of the reflective backshort.  Type-1 detectors use the invar mount as a reflective backshort, and the wafer thickness, 450~$\mu$m, sets the backshort distance to be \bst\ in silicon.  The type-2 detectors sandwich a metal film between two thin wafers to form a \bso\ backshort 150~$\mu$m behind the bolometers.  In both types, the space between the bolometers and backshort is mostly silicon.  Section~\ref{sec:perf} contains a detailed evaluation of the performance of the detectors.

\begin{figure}[th]\centering
\includegraphics[width=3.375in]{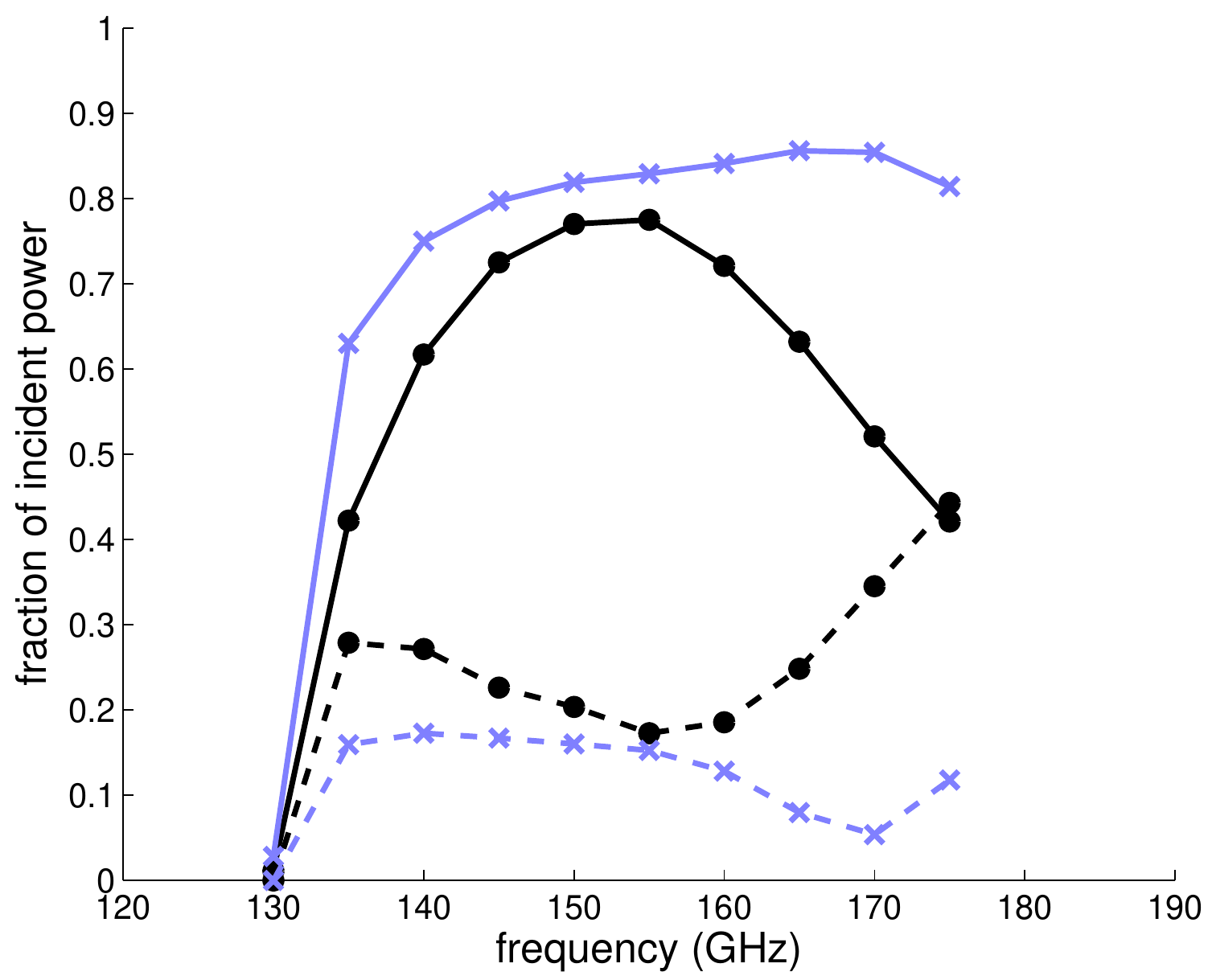}
    \caption[]{Results from HFSS simulation of the integrating cavity.  For unit power input in the horn-coupled waveguide, the plot shows power absorbed by the spiderweb (solid lines) and power radiated radially in the wafer and air gap (dashed lines) as a function of frequency.  The \bso\ backshort geometry (crosses) has better performance than the \bst\ backshort geometry (circles).}
    \label{fig:hfss}
\end{figure}

Each bolometer is centered behind a cylindrical waveguide in the horn array. Radiation not absorbed by the bolometer can leak radially to other detectors or be reflected back through the horns.  This leakage results in optical crosstalk between adjacent bolometers.  We used a 3D-electromagnetic field simulation software package, High Frequency Structure Simulator (HFSS),\footnote{Ansoft, 225 W Station Square Dr. Suite 200, Pittsburgh, PA 15219} to simulate and optimize the waveguide, cavity, and absorber geometries.  Simulated results for the \apex\ configurations are plotted in Fig.~\ref{fig:hfss}.  The \bso\ backshort has higher absorption and lower crosstalk across the observation band.  The band averaged crosstalk of the \bso\ geometry is 40\% of that for the \bst\ geometry.

\section{Detectors}
\label{sec:yield}

The \apex\ bolometers consist of an absorbing element coupled to a TES which is connected to a thermal reservoir by a weak thermal link.  The TES is a thin superconducting film with a transition temperature higher than the reservoir temperature.  A constant bias voltage is applied across the TES so the sum of electrical power dissipated in the TES plus optical power incident on the absorber heat the sensor to precisely the transition temperature.  Therefore, changes in the incident optical power produce an inverse change in electrical power.  Since the voltage bias is held constant, the change in electrical power is measured as a change in current through the sensor.

\begin{figure*}[th]\centering
\includegraphics{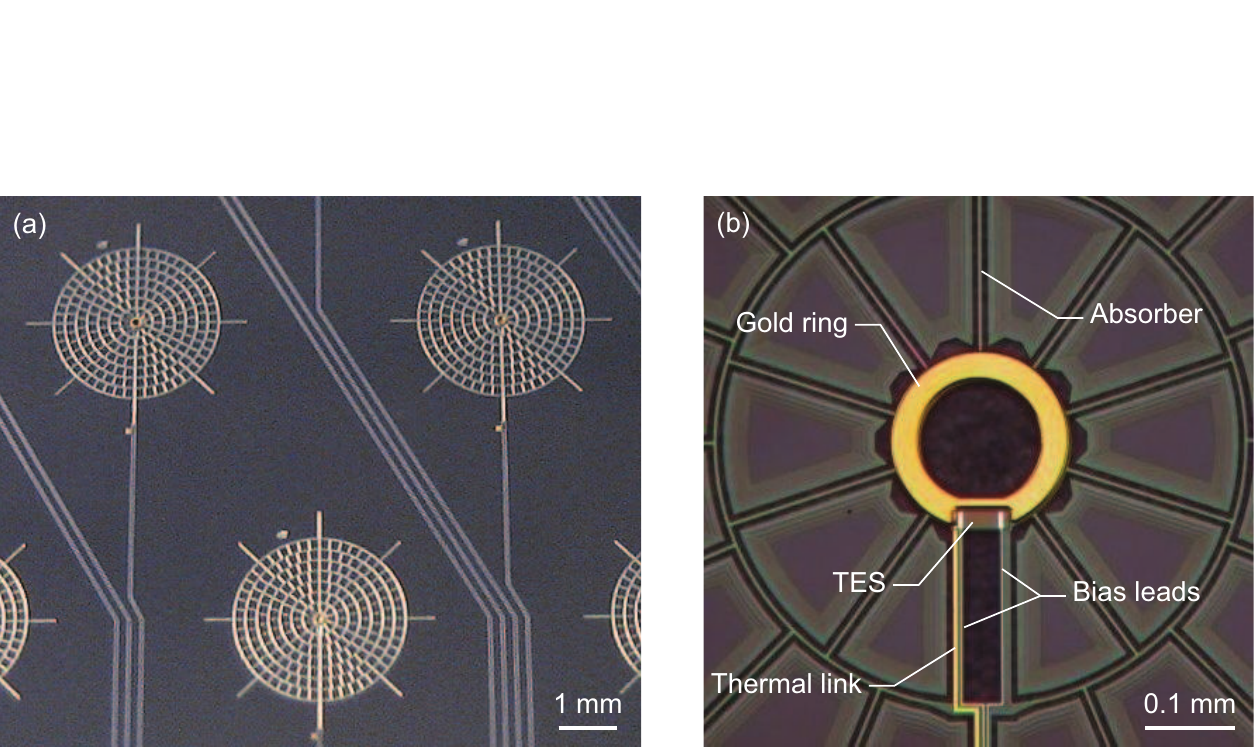}
    \caption{(a) Three spiderweb bolometers on a 55-element sub-array.  (b) A close-up of the center of the spiderweb shows the elements of the bolometer.  The dark areas along the absorber legs and inside the gold ring are regions in which the silicon has been etched away from the spiderweb.}
    \label{fig:bolo}
\end{figure*}

The bolometer absorber is a 3~mm diameter gold spiderweb, and the TES is located at its center (see Fig.~\ref{fig:bolo}). The gold thickness of the absorber is set so the spiderweb sheet resistance is 250~$\ohm/\mbox{sq}$.  This resistance is a compromise between optimal absorption, which simulations suggest peaks at 500~$\ohm/\mbox{sq}$, and fabrication constraints, which require a minimum film thickness to maintain metal continuity across the absorber.   The central location of the TES and the low heat capacity of the micromesh spiderweb absorber result in a fast optical response.   The properties of the bolometers are summarized in Table~\ref{tbl:boloprop}.

\begin{table}[th]\centering
\begin{ruledtabular}
\begin{tabular}{lcc} 
                Parameter &     Type-1 &     Type-2 \\ 
\colrule 
                  $R_{n}$ (\ohm) &        1.9 &        1.1 \\ 
                    $T_c$ (mK) &        465 &        570 \\ 
           $\overline{G}$ (pW/K) &        153 &         86 \\ 
             $\tau_{opt}$ (ms) &         11 &         20 \\ 
             Responsivity (\sci{5}\,A/W) &        1.7 &        3.0 \\ 
\end{tabular} 
\end{ruledtabular}

\caption{Typical TES Bolometer Parameters.  The terms are defined in the text of Sec.~\ref{sec:yield}.}
\label{tbl:boloprop}
\end{table}

The absorber, TES, and leads to bias the TES are deposited onto a 1~$\mu$m thick low-stress silicon nitride (LSN) spiderweb membrane.   The LSN spiderweb is suspended by eight LSN legs above a $\sim20~\mu$m vacuum gap etched in the silicon.  The gap is made by a dry XeF$_2$ gas etch, and thermally isolates the absorber from the silicon substrate.  The thermal conductance of the bolometer is set by a gold thermal link that runs from the TES to the thermal reservoir, parallel to the bias leads.  The support legs are an additional, low conductivity link to the thermal reservoir.   A fraction of the incident power absorbed on the spiderweb reaches the reservoir via the legs and does not heat the sensor as much as power reaching the reservoir via the gold link.  A finite element analysis of the thermal structure of the bolometer and the expected illumination pattern gives a calculated bolometer absorption efficiency, $\eta_{bolo}$, of 0.84 which lowers the effective efficiency of the bolometer (Sec.~\ref{sec:perf.eff}).  The thickness of the gold link is tuned so that the total thermal conductance to the reservoir, including the LSN legs, is $\overline{G}=153$~pW/K (for type-1, see Table~\ref{tbl:boloprop}).  Here, $\overline{G}$ is defined as
\begin{equation}
\overline{G} = \frac{P_{bias} + P_{opt}}{T_c - T_0},
\label{eq:powerbalance}
\end{equation}
where $T_c$ is the superconducting transition temperature is, $T_O$ is the reservoir temperature (280~mK in \apex), and $P_{bias}$ and $P_{opt}$ are, respectively, the total electrical and optical power dissipated in the bolometer.

For the detector, $T_c$ and $\overline{G}$ must be carefully chosen to avoid saturation while maximizing sensitivity.  Following equation~\eqref{eq:powerbalance}, $T_c$ and $\overline{G}$ must be large enough to dissipate the expected optical power with enough margin to allow for loading variations and sufficient bias power.  We typically use
\begin{equation}
\overline{G}(T_c - T_0) \approx 2 P_{opt}.
\label{eq:poptvg}
\end{equation}
The minimum thermal carrier noise for a bolometer where electrons are the dominant thermal carrier is given by the condition\cite{schwan.thesis}
\begin{equation}
T_c = 2.1~T_0.
\label{eq:tccond}
\end{equation}
The minimum thermal carrier noise gives the maximum contribution by photon noise to the total noise of the bolometer.  With $T_0$ fixed by our choice of cooler, equations~\eqref{eq:poptvg} and~\eqref{eq:tccond} determine the appropriate $T_c$ and $\overline{G}$ for an estimated $P_{opt}$.  In practice, achieving a specific transition temperature is difficult and optical power is not precisely known at the detector fabrication stage.  As discussed in Sec.~\ref{sec:perf.sens}, the type-2 detector parameters result in increased sensitivity relative to the type-1 detectors.

The TES is an Al-Ti bilayer with layer thicknesses tuned for a $T_c$ of 465~mK and geometry chosen for a normal resistance, $R_n$, of 1.9~\ohm (for type-1).   The TES is voltage biased and operated with strong electrothermal feedback (ETF).  The ETF maintains the TES in the superconducting transition and increases the range of the  linear response.\cite{irwin1995,*lee1996}  The loop gain of the ETF is given by
\begin{equation*}
\mathcal{L} = \frac{P_{bias}\alpha}{G T_c},
\end{equation*}
where $P_{bias}$ is the bias power and $\alpha=d\log(R)/d\log(T)$ is a measure of the sharpness of the superconducting transition.  Note that $G$ refers to the instantaneous thermal conductivity of the bolometer, $\delta P/\delta T$, which depends on the material properties of the link.
For a detector biased with a constant voltage, the current responsivity is determined by the voltage bias
\begin{equation}
S_i = \frac{\delta I}{\delta P} = \frac{-1}{V_{bias}} \frac{\mathcal{L}}{\mathcal{L}+1} \sim \frac{-1}{V_{bias}}.
\label{eq:vbresp}
\end{equation}
For the sinusoidal bias system used in \apex, the responsivity is modified by a factor $\sqrt{2}$, but the NEP for constant and alternating-biased systems are identical in the ideal case as discussed in Sec.~\ref{sec:readout}.

The ETF also speeds up the intrinsic thermal response time of the bolometer, $\tau_0=C/G$ where C is the heat capacity
of the detector, so the thermal time constant is
\begin{equation*}
\tau_{th} = \frac{\tau_0}{\mathcal{L}+1}.
\end{equation*}
This thermal time constant sets the TES response time.  In the \apex\ bolometers, the optical response time is limited by the thermalization time of the spiderweb absorber, $\tau_{opt} > \tau_{th}$.  The $\tau_{opt}$ of the type-1 and type-2 detectors is 11~ms and 20~ms, respectively.  The long time constant of the type-2 bolometers is likely due to a processing anomaly, high-heat capacity, optically-inactive residue on the spiderweb.  We must take the long time constant into account in data analysis, but the sensitivity is not affected.

For stable voltage bias, we require that the TES thermal time constant be larger than the time constant of the bias circuit.
The stability criteria given by Ref.~\onlinecite{irwin1998} is
\begin{equation*}
\tau_{\rm{th}}/5.8 \geq \tau_{\rm{bias}}.
\end{equation*}
The bias circuit time constant is dominated by the $LCR$ tank circuit of the readout described in Sec.~\ref{sec:readout} and is a function of the TES resistance, $\tau_{\rm bias}=L/R$.  The intrinsic TES thermal time constant is $< 100~\mu$s, too fast for stable operation at high ETF loop gain and much faster than the optical time constant from thermalization of the spiderweb.  Therefore, we slow the thermal response of the bolometer by adding additional heat capacity in the form of a 200~$\mu$m wide, 3~$\mu$m thick gold ring coupled to the TES.  The ring geometry is chosen to allow stable bolometer operation at loop gains of $\mathcal{L}=10\mbox{--}100$ while still keeping the bolometer thermal time constant much less than the bolometer optical time constant.

\begin{figure*}[t]\centering
\includegraphics{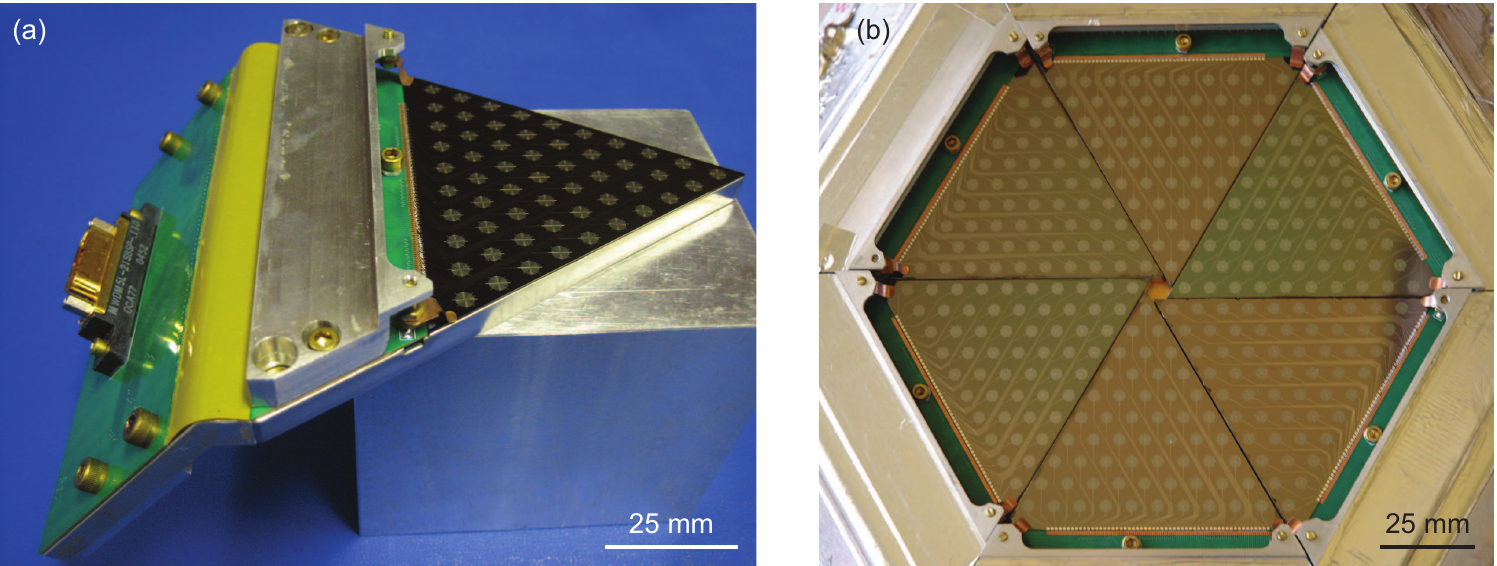}
    \caption[]{(a) A fabricated bolometer sub-array in its holder.  Each 55-element sub-array is mounted on an invar triangle and wirebonded to a multiplexer circuit board.  Multiplexer circuit elements (tuned capacitors and inductors, described in Sec.~\ref{sec:readout}) are housed under the aluminum guard.  A single micro-D connector provides the connection to the SQUIDs.  (b) The TES bolometer array.  Six identical sub-arrays are assembled into single planar array which has 330 bolometers and is 133~mm in diameter.  Each bolometer has two leads which run from the spiderweb to bonding pads at the edge of each triangle.  These are visible as light traces between the bolometers.}
    \label{fig:boloarray}
\end{figure*}

The \apex\ detector array is assembled from six identical 55-element, triangular sub-arrays (Fig.~\ref{fig:boloarray}).  Each sub-array is fabricated on a single 4 inch silicon wafer with standard optical photolithographic techniques at the Berkeley Microfabrication Laboratory.  The triangular sub-arrays are mounted on matching invar triangles with BeCu spring clips at the corners.  Invar is used to match the thermal contraction of the wafer.  Thermal contact between the wafer and invar is made by a thin film of Apiezon-N grease.\footnote{M\&I Materials Ltd., Hibernia Way, Trafford Park, Manchester M32 0ZD, UK}  The invar is itself attached to an aluminum mount which supports a multiplexer circuit board, and the bolometer sub-array is wirebonded to the board.  In the receiver, the triangular sub-assemblies form a single planar hexagonal array with 330 bolometers.  The modular design allows construction of large focal plane arrays and easy replacement of sub-arrays.

\section{Frequency Multiplexed Readout Electronics}
\label{sec:readout}

The \apex\ TES bolometers are read out with a SQUID amplifier frequency-domain multiplexing (FDM) system\cite{spieler2002,*lanting2004} which allows a number of detectors to be read out by a single 4~K SQUID amplifier connected through a single pair of wires. Readout multiplexing greatly reduces the thermal load on the low-temperature stage, the complexity of cold wiring, and the system cost. This technique enables the use of large detector arrays.  The \apex\ system uses seven detectors read out by one SQUID amplifier in each multiplexer module.

\begin{figure*}[t]\centering
\includegraphics{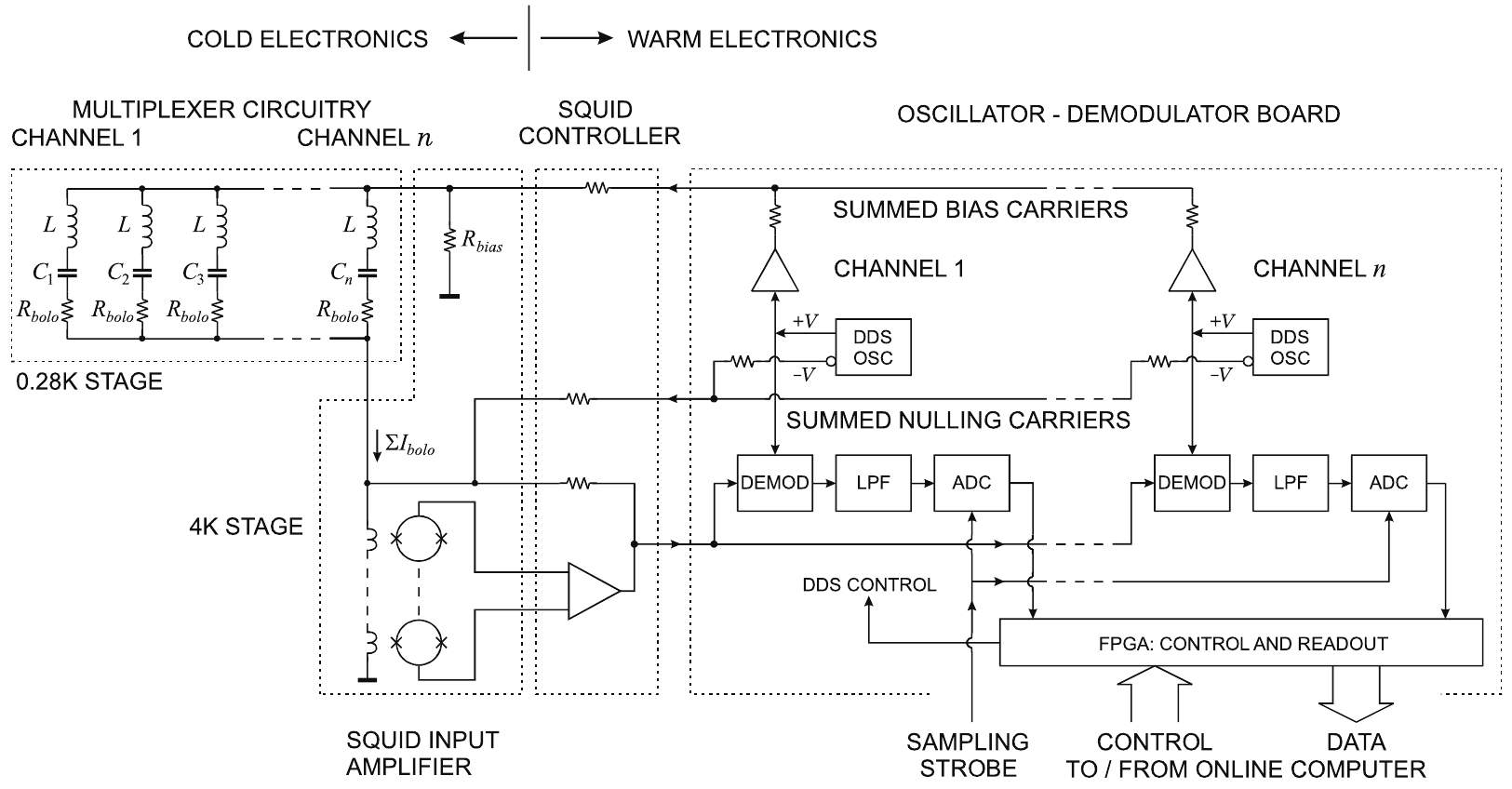}
    \caption[]{Schematic diagram showing a single module of the frequency multiplexed SQUID readout system. Sinusoidal bias voltages at different frequencies are summed and applied to a group of bolometers via a single line.  $LC$ filters in series with the bolometers isolate an individual bias for each bolometer.  The modulated signals from the bolometers are summed at the input of a SQUID array.}
    \label{fig:muxBlockDiagram}
\end{figure*}

The basic components of a readout system module are shown in Fig.~\ref{fig:muxBlockDiagram}. The bolometers are biased with alternating voltages at carrier frequencies (0.3--1 MHz) that are much higher than the bolometer thermal bandwidth, so the bias deposits a power on the sensor that depends only on its resistance. Each bolometer in the module is biased at a different frequency. Changes in absorbed radiation change the resistance of a given bolometer and modulate the current. This amplitude modulation translates the absorbed signal spectrum to sidebands centered around the carrier bias frequency. Since the currents from the different bolometers are separated in frequency, they can be combined and transmitted to the SQUID amplifier with a single wire. Furthermore, the bolometers are connected through series $LC$ circuits, each of which is tuned to the appropriate bias frequency, so all bias frequencies can also be fed through a single line. As a result, each multiplexer module requires only a single pair of wires from the low-temperature stage to the 4~K stage to read out all the bolometers.  The comb of amplitude modulated carriers is fed to a bank of demodulators that mix the signals down to base-band. The signals are then filtered and digitized. All outputs in the array are sampled synchronously at 1~kHz, a much higher frequency than the signal bandwidth. The data streams are passed though a digital low-pass filter,  downsampled to 100~Hz, and compressed before being written to disk by the readout computer.

High impedance NTD bolometers are susceptible to spurious signals induced by microphonic vibrations\cite{richards1994} of the bolometer wiring.  However, TES bolometers are low impedance devices which are relatively insensitive to vibrations.  Using an alternating voltage bias further reduces the sensitivity to microphonics.  Since the detectors only respond to voltages near their bias frequency, the FDM system is insensitive to microphonic excitation of the bolometer wiring which occurs at much lower frequencies.  Thus, no special effort to restrain the bolometer wiring is required for this system.

The bandwidth of the $LC$ circuits is chosen to accommodate the signal sidebands and attenuate the crosstalk between neighboring channels to an acceptable level ($<1\%$). The bandwidth is the same for all channels and is determined by the inductance and the bolometer resistance, $R_{bolo}/(2\pi L)$.  The high-Q $LC$ filters are composed of photolithographically constructed 16~$\mu$H spiral inductors and commercial negative-positive-zero (NP0) ceramic chip capacitors that are mounted on a printed circuit board adjacent to the bolometer sub-array.  The resulting filter is compact and comparable in area to a single bolometer. In addition to defining the carrier frequency for each channel, the tuned circuits also limit the bandwidth of the bolometer Johnson noise, which would otherwise contribute to the noise in all other channels of the module.

The SQUID amplifier system has been carefully designed to achieve the required low input impedance, low noise, and high dynamic range across the MHz carrier bandwidth.   Low input impedance is necessary  because voltage biasing requires that the impedance of the SQUID preampliflier be much smaller than the bolometer resistance.  Shunt feedback is applied to further reduce the input impedance of the preamplifier.  To achieve the necessary transimpedance and dynamic range, an integrated array of 100 SQUIDs with inputs and outputs connected in series\cite{welty1991,*huber2001} is employed. The SQUID arrays were designed and fabricated at NIST.\footnote{National Institute of Standards and Technology}  The array allows a smaller total input inductance and increased signal-to-noise relative to a single SQUID.  A feedback loop gain of about 20 is achieved by feeding the SQUID output to a room-temperature amplifier with low noise and a sufficiently large bandwidth to maintain stability over the full range of bias frequencies.

The bolometer signal information appears in the sidebands of the carrier frequency while the amplitude of the carrier contains no useful information.   Thus, the dynamic range requirements of the SQUID system can be reduced by nulling the carrier without losing any signal information.  At each carrier frequency, we add a constant amplitude nulling current which is adjusted to be out of phase by $\pi$ at the SQUID input.  The resulting nulling of the carrier reduces the total signal amplitude through the SQUID amplifier system by more than a factor of 10.  The bias and nulling signals take different paths through the cryogenic system and so they incur different phase shifts due to inductive and capacitive strays such as the wiring inductance and the reactance of the $LC$ filters on neighboring channels.

To tune the phase of the nuller waveforms efficiently it is necessary to measure the magnitude of the residual carrier. This measurement is achieved by using an eighth multiplexer channel from each module to measure the magnitude of the residual bias for the other channels during the nulling process. This strategy improves the speed and accuracy of the carrier nulling at the expense of a channel which would otherwise be used to readout an additional detector.  Each multiplexer module has eight channels, seven used for bolometer readout and one used for nulling.

\begin{figure*}[t]
\centering
\includegraphics{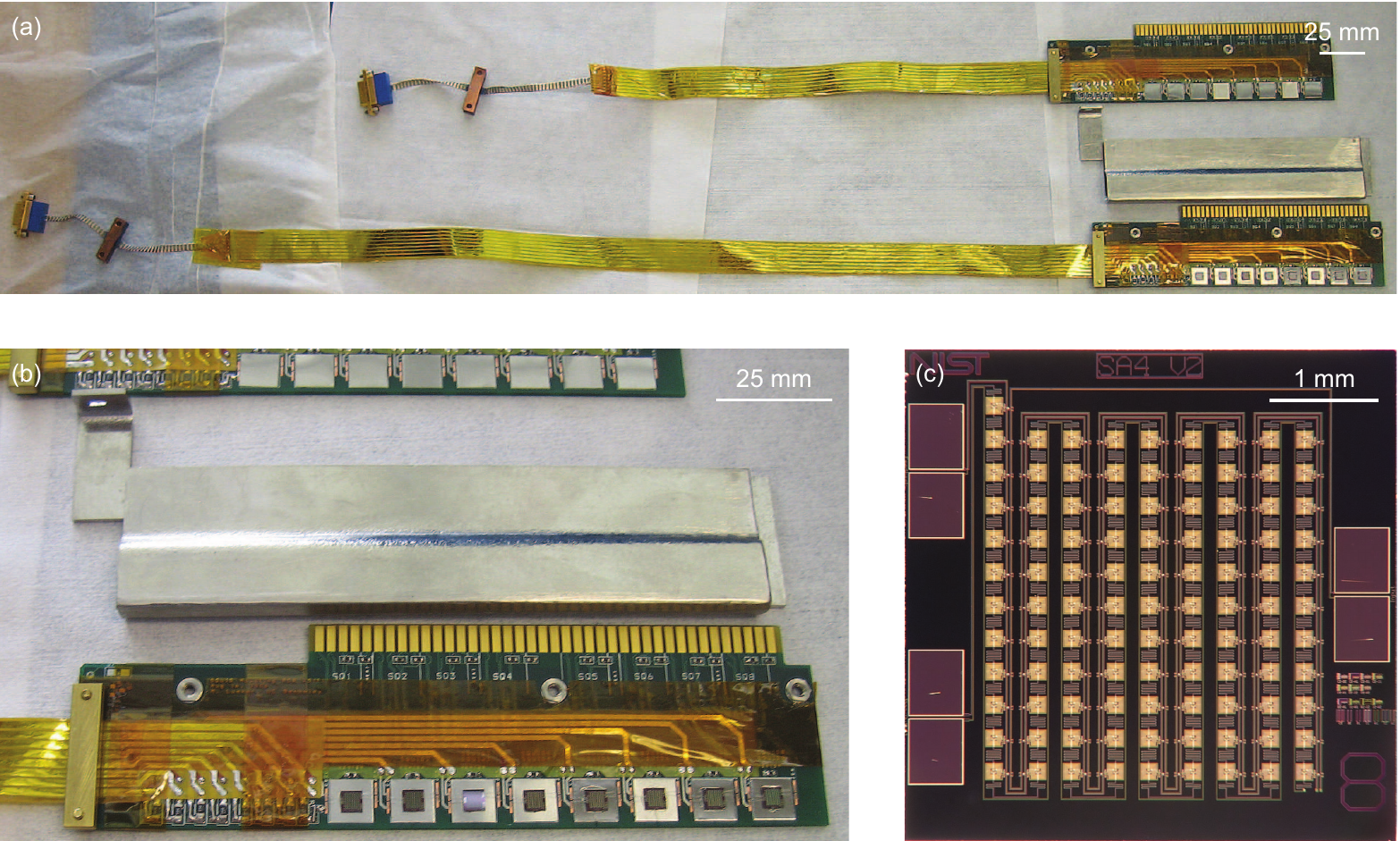}
\caption[]{(a) Two SQUID cards with 4~K to 280~mK wiring and a Cryoperm sheath.  Traces on flexible kapton ribbon, soldered directly to the SQUID card, are in series with woven NbTi cable.  In \apex\ we use 510 and 360~mm lengths of flexible-ribbon cable.  The NbTi cable has a copper block epoxied at its midpoint for heatsinking at 370~mK. The male micro-D connector at the end of the cable mates with the female connector on the multiplexer board shown in Fig.~\ref{fig:boloarray}. (b) A detailed image of the SQUID card with magnetic shielding.  The SQUID card has 8 SQUID arrays, each of which is mounted on a square pad of niobium foil.  The card slides into the Cryoperm sheath shown above the card, leaving the gold contacts exposed for inserting into a Peripheral Component Interconnect (PCI) connector.  The left side of the card has 8 bias chip resistors.  The upper edge of the photograph shows part of a SQUID card before the SQUID arrays have been attached.  (c) A close-up image of the 100-element series array SQUID chip.}
    \label{fig:microstrip}
\end{figure*}

The seven bolometers per module are read out by a single SQUID amplifier. For each triangular sub-array of bolometers, seven groups of seven bolometers are each fed to a single SQUID card which holds the seven SQUID arrays.  Thus, the readout requires fourteen wires for each sub-array.   Note that there happen to be seven bolometers for each SQUID array and seven readout SQUID arrays on each card, but these two factors are independent.  The wiring must have low thermal conductance to minimize the load on the millikelvin stages, and low impedance to minimize shifting of the $LC$ resonance frequency.  For this purpose, we use a hybrid wiring shown in Fig.~\ref{fig:microstrip}: low inductance copper traces coated with tin-lead solder on flexible kapton in series with low thermal conductance NbTi twisted-pairs.  The traces are 1.8~mm wide, deposited on both sides of a 18~mm wide kapton ribbon, thereby limiting the inductance to 0.05~nH/mm.  The NbTi wires have a CuNi coating for soldering, are woven into a cable,\footnote{Tekdata, Ltd, http://www.cryoconnect.com/} and have 1~nH/mm inductance.  Both parts of the hybrid cable are superconducting.  Using 510~mm of flexible-ribbon cable with 100~mm of twisted NbTi cable provides an acceptable compromise between inductance, $\sim150$~nH total, and thermal load (Sec.~\ref{sec:mkstage}).

In the \apex\ configuration, we use six SQUID cards each with seven SQUID arrays to read out bolometers.  Each SQUID card also has an eighth SQUID array which is not wired to bolometers and can be used for readout diagnostics.  The SQUIDs are extremely sensitive to magnetic fields and must be carefully shielded. Each SQUID is mounted on a 9~mm square pad of superconducting niobium foil (Fig.~\ref{fig:microstrip}), which serves to trap residual flux during cool down and attenuate time-varying magnetic fields.  Further, the SQUID card is housed inside its own Cryoperm\footnote{Vacuumschmelze GmbH, Gruner Weg 37 D-63450, Hanau, Germany} magnetic sheath.  Cryoperm is an alloy which has high permeability at cryogenic temperatures.  The relative permeability at 4~K is $70,000$ for 0.4~A/m static fields, comparable to the permeability of mu-metal at room temperature.  The Cryoperm shield attenuates both time-varying and spatially varying fields.  We measure a reduction of field variations by a factor 2.5\sci{4} with this two-fold shielding scheme.  At this shielding level, SQUID movement through the Earth's magnetic field is detectable at signal-to-noise ratio of near unity. However, such low spatial frequencies are filtered by the analysis pipeline.


As noted above, the high gain-bandwidth product of the feedback loop requires short connections between the 4~K SQUID arrays and room temperature op-amps to minimize phase shifts and maintain stability of the negative feedback loop at megahertz frequencies and high loop gains.  The amplifiers are mounted on ``SQUID controller'' circuit boards which are attached directly to the side of the receiver cryostat wall in an RF-tight aluminum box.  There is a one-to-one correspondence between SQUID boards and SQUID controllers; each SQUID controller board houses the op-amps and control logic to provide bias currents for the eight SQUID-arrays on each SQUID board.  To couple the 4~K SQUID boards to the room temperature SQUID controllers, we use a custom manufactured\footnote{Tekdata} wire harness consisting of 518 120~mm length manganin wires in woven Nomex.  Up to seven SQUID cards can be plugged into a single wiring harness.

The comb of sky-signal modulated carriers is passed from the SQUID controller to a bank of demodulators on one of twenty 16-channel oscillator-demodulator boards. Direct Digital Synthesizers (DDSs) provide carrier and nulling signals with precise frequency and amplitude control and very low sideband noise.  The same DDS that generates the bolometer bias also provides a phase and frequency locked square-wave reference for the corresponding demodulator. In the absence of phase shifts in the cryostat, the demodulator output would correspond to the in-phase $I$-component of the carrier signal. However, the wiring strays result in a small offset in phase between the demodulator and carrier that increases with frequency. The lack of phase adjustment in the demodulator means this phase contributes to a slight degradation of noise performance at high carrier frequency (see Sec.~\ref{sec:perf.sens}).
The demodulator circuits use a sampling demodulator followed by an 8-pole, low-pass anti-aliasing filter, and a 14-bit analog-to-digital converter (ADC).  A field programmable gate array (FPGA) assembles the data from all channels on a given board and streams it to the data acquisition computer.
Each demodulator board can process the outputs from two SQUID amplifiers, i.e. two groups of eight multiplexer channels. The readout crate can accommodate twenty demodulator boards with a total of 320 channels.  However, since only seven channels in each group are used to read bolometers, the maximum number of bolometer channels is 280.  There are seven channels for each group of eight bolometers.

\section{Cryogenics}
\label{sec:cryogenics}

\subsection{Cryostat}
\label{sec:cryostat}

The \apex\ cryostat houses cold filters and lenses, the detector array, and SQUID electronics.  The focal plane is cooled by a 3-stage helium sorption refrigerator backed by a mechanical pulse-tube cooler; Table~\ref{tbl:coolpower} summarizes the available cooling power at various temperature stages.  The mechanical cooler eliminates the need for open reservoirs of liquid cryogens, greatly simplifying the design and construction of the cryostat while reducing the cost and difficulty of remote observations.  The receiver is essentially downward-looking, further simplifying the cryostat design. The optical and pulse-tube axes are parallel while maintaining the pulse-tube cooler within 30\dg\ of vertical orientation as required for its operation.

\begin{table}[th]\centering
\begin{ruledtabular}
\begin{tabular}{lc}
  Stage& \dcoltext{Cooling Power (W)} \\
\colrule
  60~K & 40\\
  4~K & 1\\
  350~mK & 6.0\sci{-5}\\
  270~mK & 1.5\sci{-6}\\
\end{tabular}
\end{ruledtabular}
\caption{Cooling power provided by the pulse-tube cooler and the 3-stage helium sorption refrigerator at different temperature stages of the \apex\ cryostat.}
\label{tbl:coolpower}
\end{table}

\begin{figure*}[t]\centering
\includegraphics{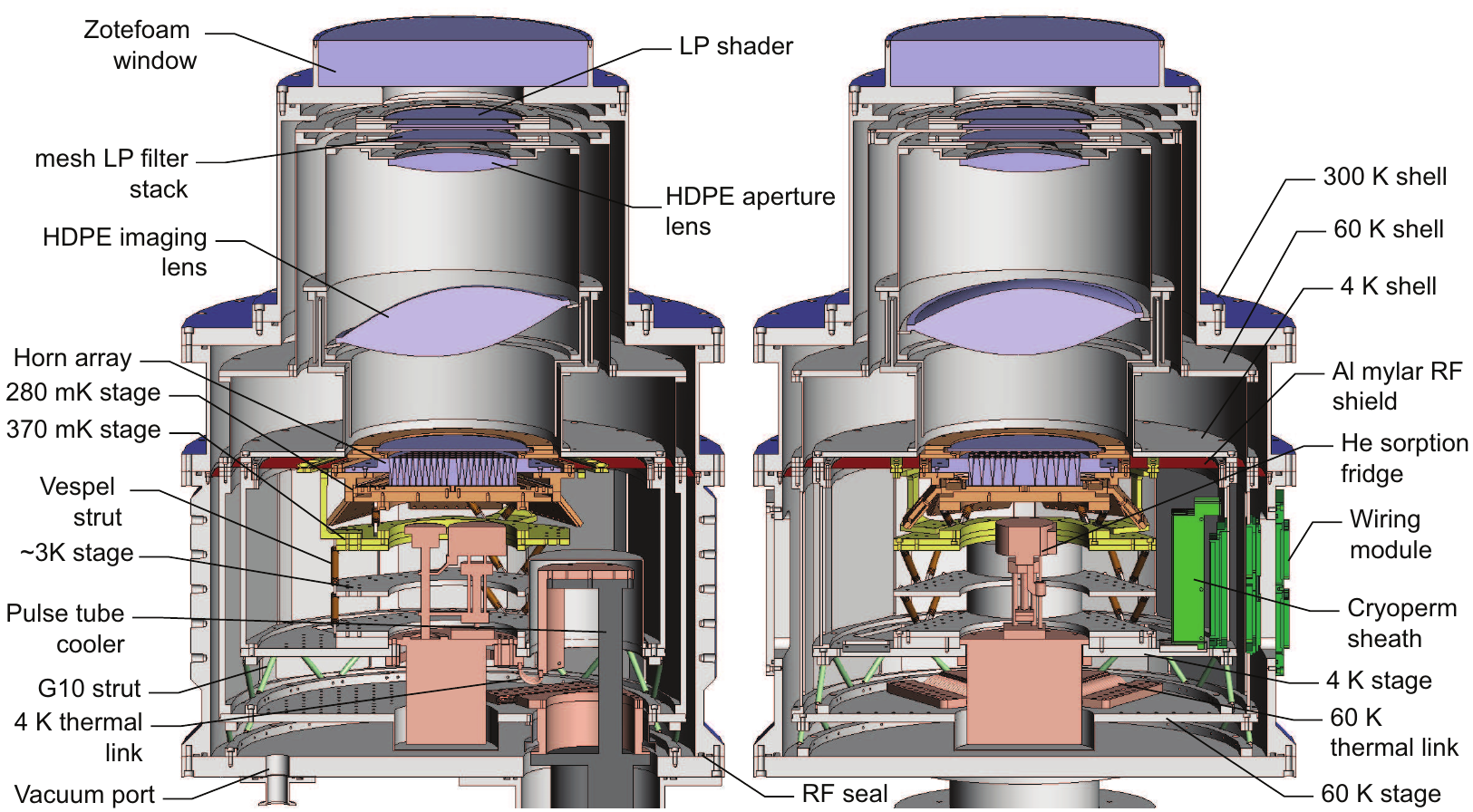}
    \caption[]{Drawings showing two orthogonal sections of the \apex\ cryostat.  Many components are labeled.  The cryostat is shown looking upward in the orientation used in the laboratory for work on the focal plane. The optical section (upper half) can be removed for access to the focal plane and readout components.  The aluminized mylar sheet (red) acts an RF shield and separates the optical and readout sections of the cryostat.  Optical components are shown in light blue, 280~mK components in orange, 380~mK in yellow, and the 4~K SQUID card housing and readout wiring module in green.  When mounted in APEX, the cryostat is downward-looking, with the window pointed at a mirror on the floor of the cabin.}
    \label{fig:cryostat}
\end{figure*}

Cutaway drawings of the inverted cryostat are shown in Fig.~\ref{fig:cryostat}.  The cryostat includes 4~K and 60~K radiation shields supported by G-10 fiberglass struts inside the 300~K vacuum shell.  The 60~K radiation shield is gold plated to reduce its emissivity and decrease the radiation load on both the 4~K and 60~K shields.  Where feasible, we wrap cylindrical portions of both shells with aluminized mylar to further reduce the radiative load.   The optical section can be removed for access to the focal plane and SQUID electronics. The section of the receiver which houses the wiring modules was designed to be modular, allowing aspects of the cryostat design to be used for future experiments.

Because the APEX telescope hosts a number of instruments, the uncertain RF environment of the cabin was a major concern.  The lower section of the cryostat (as oriented in Fig.~\ref{fig:cryostat}) is an RF sealed enclosure.  All vacuum o-ring seals also include an additional RF seal of elastofoam.\footnote{Tecknit USA, 135 Bryant St, Cranford, NJ 07016}  SQUID controller cards are housed in an RF tight box mounted on the cryostat, with boards plugging directly into the hermetic feedthroughs on the cryostat shell.  Commercial pi-filter connectors on the outer face of the box provide feedthroughs for the readout wiring.  Inside the cryostat, aluminized mylar sheets create electrical shorts between the 300~K, 60~K and 4~K shields.  An additional sheet between 4~K and 280~mK completes the shielded compartment.  The only perforations are the feed horn array, hollow screws to connect vacuum spaces, and the electrical feedthroughs on the room-temperature electronics box. The feed horns are coupled to waveguides with a 130~GHz cutoff frequency, the holes in the hollow screws have a high length to diameter ratio, and the
electrical feedthroughs are protected by pi-filters.  The thickness of the aluminum layer on the mylar is a compromise between electrical and thermal resistance, with better shielding also increasing the thermal load on low temperature stages.  We take advantage of the larger cooling power available at higher temperatures and tailor the  aluminum thickness for each temperature stage: 400~nm for the 300--60~K temperature gap, 100~nm for 60--4~K, and 50~nm for 4--0.280~K.

Heat straps from the cryostat cooling elements---the pulse-tube cooler heads and sorption refrigerator millikelvin stages---to the cryostat are all made from lengths of 12~AWG\footnote{equivalent to 2.05~mm diameter solid wire} braided oxygen-free high conductivity (OFHC) copper wire melted into solid copper blocks at each end by tungsten inert gas (TIG) welding.  Braided wire was chosen to decouple the vibrations of the pulse-tube from the receiver and allow differential thermal contraction between the cryostat components.
The solid copper ends give high thermal conductance interfaces.

\subsection{Pulse-Tube Cooler}
\label{sec:ptc}

\apex\ uses a commercial pulse-tube refrigerator, model PT410-OP3 manufactured by Cryomech, Inc.,\footnote{13 Falso Dr., Syracuse, NY 13211} to provide cooling at 4~K and 60~K.  In a pulse-tube cooler, high pressure He gas from a compressor is passed through a high heat capacity regenerator and then allowed to expand while in contact with a cold head, absorbing heat from the head.  A motor outside the cryostat drives a valve which controls the pressure cycle of the pulse-tube.  There are no moving parts or valves in the cold section.  This feature makes the pulse-tube cooler lower vibration, more robust, and easier to maintain than other mechanical coolers.

The pulse-tube cooler has a number of advantages over liquid cryogens for remote observations, but it can add noise to measurements via several mechanisms: electrical noise, mechanical vibration, and temperature oscillations.  The compressor and valve motor are electrically noisy and can cause oscillating voltages on the cryostat shells.  The cooler used in \apex\ has its valve motor separated from the pulse-tube by a 500~mm length of flexible tubing including a short section of HDPE which electrically isolates the compressor and valve motor from the cryostat.  In addition to electrically isolating the valve motor, we drive it with a low-noise linear stepper motor.\footnote{Precision Motion Controls, 160 E. Virginia St. \#264, San Jose, CA 95112}  Valve motor movement and gas flow propagate vibrations along the pulse-tube and cause the cold head to vibrate.  The pulse-tube is mechanically isolated by mounting it on an elastomer vibration damper.\footnote{National Electrostatics Corp., 7540 Graber Road, P.O. Box 620310, Middleton, WI 53562}  Flexible copper braids isolate components inside the cryostat from vibrations on the cold heads.  The temperature of the cold head oscillates with the pressure cycle.  The bare pulse-tube cooler head has 200~mK temperature oscillations at the 1.4~Hz cycle frequency but these are attenuated by the large heat capacity of the cryostat.  The resulting oscillations at different cryostat stages are detailed in Sec.~\ref{sec:thermometry}.

An additional concern when using the pulse-tube cooler rather than liquid cryogens is that the cooling performance varies with the orientation of the pulse-tube relative to gravity.  The best performance is achieved with the pulse-tube pointing downward.  The PT410 loses about 10\% of its cooling power when tilted 30\dg\ from vertical, with more severe losses at larger angles.  The tertiary optics, cryostat and mount were designed to hold the cryostat and pulse-tube at 30\dg\ when pointed at zenith.  Thus, for observations between zenith and 30\dg\ elevation, the pulse-tube is tilted at less than 30\dg\ relative to vertical.

\subsection{Helium Sorption Refrigerator}

We use a custom built 3-stage helium sorption refrigerator from Chase Research\footnote{140 Manchester Road, Sheffield, S10 5DL, UK} to cool the bolometers to 280~mK. It is sometimes referred to as a ``He-10'' refrigerator because it includes one \hef\ stage and two \het\ stages.    The design includes sintered copper in the helium reservoirs to provide good thermal contact between the fluid helium and the refrigerator cold heads over a range of orientations.  Many previous bolometric CMB experiments such as the Arcminute Cosmology Bolometer Array Receiver (ACBAR; Ref.~\onlinecite{runyan2003} and Bolocam have used similar refrigerators backed by open reservoirs of liquid cryogens.

A nominal He-10 refrigerator cycle cools the \apex\ focal plane to 280~mK for more than 24~h.  The \hef\ is cycled only to condense \het\ into the ``Interhead'' and ``Ultrahead'' reservoirs  which cool to 350~mK and 270~mK, respectively.  At these temperatures (and ignoring parasitic loads), the Interhead has 60~$\mu$W of cooling power, absorbing 5~J during the cycle. It serves to buffer the Ultrahead, which has 1.5~$\mu$W of cooling power, absorbing 0.1~J during the cycle.  Heat from the wiring, RF shielding and focal plane support structure is intercepted at the Interhead.

The efficiency of the refrigerator cycle as a whole depends critically on the temperature of the condensation point of the \hef\ stage.  It is therefore thermally connected directly to the 4~K head of the pulse-tube via braided cooper wire as described in Sec.~\ref{sec:cryostat}.

The He-10 refrigerator is cycled by applying currents to six resistors.  Three operate the charcoal pumps, and three operate gas-gap heat switches between the pumps and 4~K head.  The switches are closed by using a resistor to heat a small charcoal pump and release helium into the space between the hot and cold surfaces of the switch.    An electronics box with computer control logs temperatures (Sec.~\ref{sec:thermometry}) and allows remote, automatic cycling of the refrigerator in 2.5~h.  At the end of the cycle, we bias the bolometers for observation while they are in the normal state and then allow the focal plane to cool below $T_c$ with the bolometers biased.

\subsection{Millikelvin Stage}
\label{sec:mkstage}

The support structure for the focal plane provides a rigid mechanical support while maintaining the thermal isolation required to achieve millikelvin temperatures.   As shown in Fig.~\ref{fig:cryostat}, the focal plane support consists of three stages supported from the 4~K stage.  The structure provides temperature intercepts with each stage thermally anchored to a different point of the He-10 refrigerator: the $\sim3$~K stage is thermally connected to the ``Heat Exchanger'' of the refrigerator (a point which equilibrates between the 4~K stage and Interhead temperatures); the $\sim370$~mK stage is thermally connected to the Interhead, and the $\sim280$~mK stage is thermally connected to the refrigerator Ultrahead.  The focal plane and $LC$ filter boards are bolted to the $\sim280$~mK stage.  Wires to the focal plane and the aluminized mylar RF shield are thermally anchored at the $\sim370$~mK stage.  The NbTi section of low-inductance hybrid wiring (described in Sec.~\ref{sec:readout}) is thermally anchored at the $\sim370$~mK stage resulting in loads of 16~$\mu$W on the Interhead and 0.03~$\mu$W on the Ultrahead.

The stages are made from gold-plated aluminum, and each level is supported from the one below it by six Vespel\footnote{Dupont, www.dupont.com} legs.  We use Vespel SP-1 for supporting the $\sim3$~K and $\sim370$~mK stages.  The load from the supports on the Interhead is 19~$\mu$W.  We use graphite loaded Vespel SP-22 to support the $\sim280$~mK stage due to its lower thermal conductance compared to unloaded Vespel at millikelvin temperatures.\cite{locatelli1976}  The load on the Ultrahead from the supports is 0.3~$\mu$W.  The full thermal loading on the two heads is summarized in Table~\ref{tbl:mkloads}.

\begin{table}[th]\centering
\begin{ruledtabular}
\begin{tabular}{lcc}
   & Interhead ($\mu$W) & Ultrahead ($\mu$W) \\
\colrule
  Support structure & 19 & 0.32 \\
  RF Shielding & 15 & 0.15 \\
  Wiring & 16 & 0.03 \\
  Radiation & 0.004 & 0.19 \\
  Bolometers & \ldots & 0.01 \\
  \\
  Total & 50 & 0.70\\
\end{tabular}
\end{ruledtabular}
\caption{Power Dissipation on mK Stages}
\label{tbl:mkloads}
\end{table}

\subsection{Thermometry}
\label{sec:thermometry}

\apex\ uses commercial thermometers\footnote{Lakeshore Cryotronics, 575 McKorcle Blvd, Westerville, OH 43082} read out by the data acquisition computer to monitor temperatures inside the cryostat.  The refrigerator millikelvin heads and their corresponding stages are monitored with Cernox RTD thermometers, accurate from 300~K to below 280~mK and read with a AC bridge.
Silicon diode thermometers monitor refrigerator pumps and heat switches as well as both the 60~K and 4~K pulse-tube heads
and corresponding stages. The temperatures returned from the thermometers as well as the states of the six heaters on the sorption refrigerator are all continuously monitored via an electronics box in the Cassegrain cabin similar to that used in ACBAR.

We do not use active temperature stabilization at 4~K or on the focal plane.  Slow temperature drifts of the focal plane or the optics have an observable effect on the bolometer signal, but are well below the signal band.  The  pulse-tube cooler has 1.4~Hz temperature oscillations which cause 2~mK fluctuations at the 4~K cryostat stage.  The temperature fluctuations at the Ultrahead are $<20$~nK.  These fluctuations induce a $\sim 2~$aW signal in the bolometer which is well below the detector noise level.  Temperature fluctuations in the cold optics cause variable loading on the bolometers, but these are damped by the thermal resistance along the radiation shells and the heat capacity of the optics.   The Lyot stop has temperature fluctuations $<3~\mu$K, roughly equivalent to a $3~\mu$K CMB temperature fluctuation which is well below the detector noise.  Because the scan strategy is asynchronous with pulse-tube fluctuations, any resulting signal would appear as an insignificant additional noise in the final maps.

\section{System Performance}
\label{sec:perf}

The \apex\ receiver was first installed at the APEX telescope in December 2005 for an engineering run with a single 55-element sub-array.\cite{dobbs2006}  We redeployed the system with a full 280 channel array of type-1 detectors in early 2007.  In 2009, we replaced one type-1 sub-array with a type-2 sub-array.

As a millimeter-wave instrument that is nominally less affected by
atmospheric water vapor than shorter wavelength receivers, \apex\ has
observed in a range of weather conditions.  In bad weather, detector sensitivity can degrade by up to a factor two, and aggressive timestream data filtering is required to eliminate large spatial-scale noise due to temporally varying water vapor fluctuations. A full characterization of the instrument and the Atacama site in various weather conditions is beyond the scope of
this paper.  Therefore, as an accurate assessment of the receiver's
capabilities, we analyze data taken during excellent weather at the
height of the Austral winter.  In this section, we use data from an
August 2007 observing period to characterize the optical performance
of the system and the sensitivity of the type-1 detectors.  The type-2
detectors are characterized using data from April 2009.

\subsection{Detector Yield}
\label{sec:perf.detect}

Typically, 170--180 of the 280 type-1 channels in the receiver were optically responsive.  One of the sub-arrays, fabricated earlier than the others,  was unstable and had no live bolometers (as indicated by the wedge-shaped
void in Fig.~\ref{fig:marsbeams}).  Two SQUIDs had anomalously high noise, which resulted in the loss of 14 channels.  In the other five sub-arrays, fabricated with a more mature process, 90\% of the TESs could be biased in their transition with strong ETF.

The gold layer on the spider-web absorber is as thin as the manufacturing process allows to match the characteristic wave impedance in the cavity.    It is likely several bolometers have incomplete gold coverage which reduces optical absorption.  About 85\% of the bolometers which can be stably biased are optically sensitive enough to map bright calibration sources.  However, noise filtering results in removing additional noisy and low sensitivity channels.  After data processing, there are typically 120--140 type-1 channels which contribute to a cluster map.

On the type-2 sub-array installed in 2009 and used in this analysis, 70\% of the bolometers can be stably biased in the transition. After data processing, there are typically twenty type-2 detectors which contribute to a map.

\subsection{Beams}
\label{sec:perf.beams}

The \apex\ beams are measured with a raster scan of a planet.
The rasters are \skypatch{0.5}{0.5} with azimuthal sweeps separated by 18\asec\ in elevation.   This scan pattern is wide enough to sample the full array and dense enough to ensure that each beam is fully sampled.

\subsubsection{Gaussian Fit}
\label{sec:beam.gauss}

For each channel, timestream data are binned into an individual beam map.  The planet source is frequently extended in our beam, therefore, each map is fit to the convolution product of a 2-D Gaussian and a planet-sized disk.  The resulting best-fit 2-D Gaussian determines the size, ellipticity and pointing offset for each beam.

\begin{figure}[th]\centering
\includegraphics[width=3.375in]{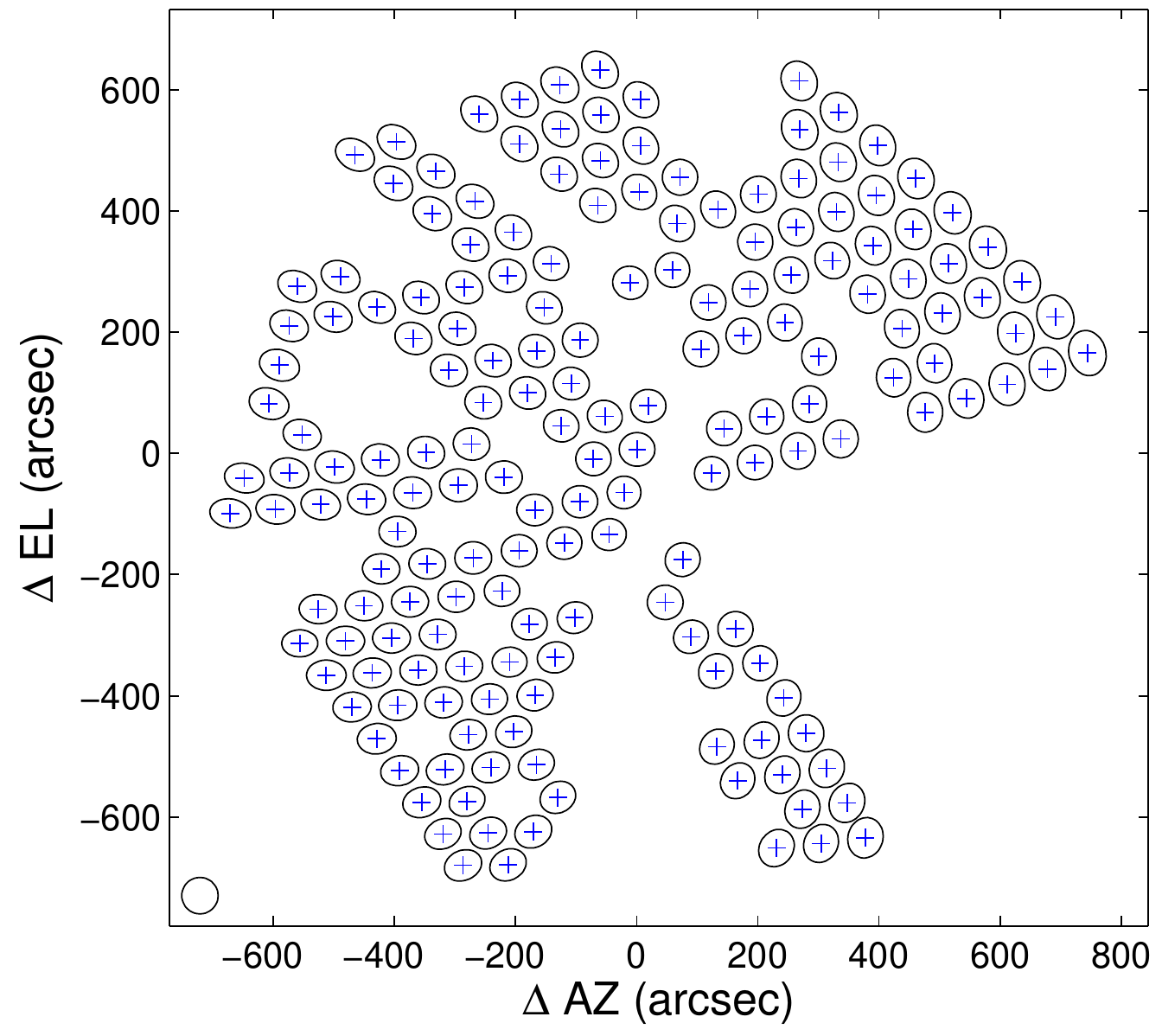}
    \caption[]{Elliptical Gaussian beam fits of 177 optically live channels.  Ellipses represent best fit 2-D Gaussian FWHM of individual channel maps; pluses mark the center of each beam.  The circle in the lower left corner is 60\asec.  The ellipticity and orientation of the beams change slightly across the array.  The median beamwidth is $58\pm 6\asec$ with a median axial ratio of 1.17.}
    \label{fig:marsbeams}
\end{figure}

\begin{figure}[th]\centering
\includegraphics[width=3.375in]{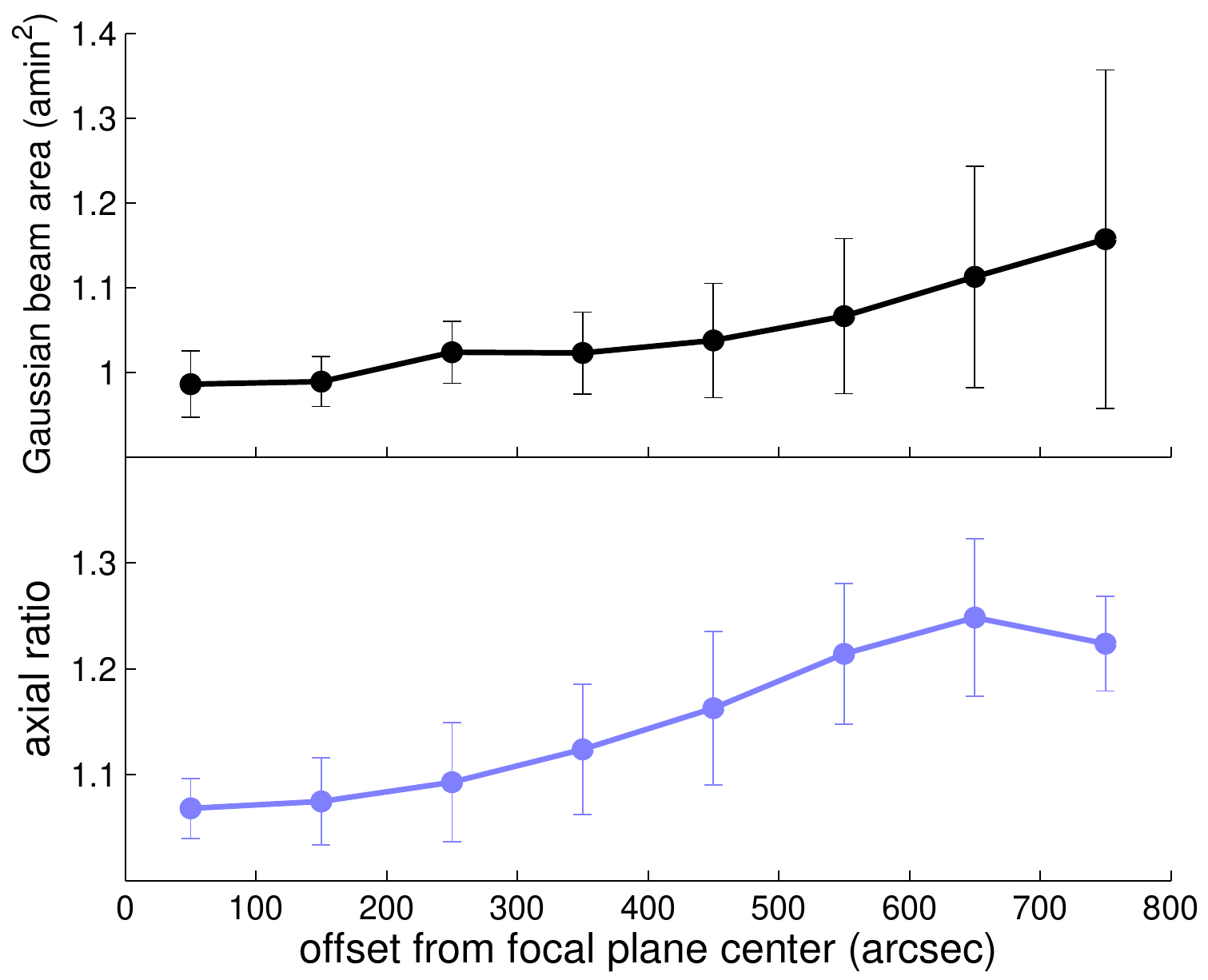}
    \caption[]{The lines show the mean Gaussian beam area (black) and axial ratio (blue) versus radial offset from the focal plane center. Error bars show the standard deviation of each value for the pixels in each radial bin.  The beam area and axial ratio increase by $17\%$ from the center to edge of the array, but the variation in beam area increases by a factor five.}
    \label{fig:fwhm}
\end{figure}

Figures~\ref{fig:marsbeams} and~\ref{fig:fwhm} show the result from a typical Mars scan.  During this scan, the angular size of Mars was 8\asec.  The median FWHM beamwidth is $58\pm 6\asec$ with a range of 45--75\asec\ and a median axial ratio of 1.17. The median beam area is $1.0\pm 0.1~\sqamin$.  The measured FWHM beamwidths and axial ratios agree quite well with those expected from the optics design.  A sample of nine beams across the array modeled using \textsc{zemax} physical optics propagation have median FWHM beamwidth of 57\asec\ with range of 47--75\asec\ and median axial ratio of 1.09.  The beam size, ellipticity and orientation change slightly across the array, with ellipticity increasing toward the edges.

The size and orientation of the outer beams is sensitive to the focus position of the secondary mirror.  The best focus position is that which minimizes the median beam FWHM across the array.  This position is slightly offset from the best focus position for a channel near the array center.  We typically measure this offset distance at the beginning of an observation period by making a set of beam maps at various focus positions.  For subsequent days, we find the best focus position for a single central pixel and apply the secondary offset.  The offset is remarkably consistent between observing periods.

\subsubsection{Beam Profile}
\label{sec:beam.prof}

In addition to a central Gaussian lobe, each beam has an extended non-Gaussian profile.  Mapping the extended profile for each beam would require a prohibitively long scan, over 50~h of integration on a calibration source.  Therefore, we create a coadded map of the average beam by combining individual beam maps with minimum variance weighting, using the sample variance of the timestream data.  Though the individual beams are elliptical and their orientation varies, the composite map is very nearly radially symmetric.  We take the radial average to derive a single high signal-to-noise beam profile.

The measured mean beam profiles for both detector types have a central lobe that is best fit by a 59\asec\ Gaussian.  The profiles differ in the level of the near sidelobes, $-14$ and \db{-15} for the type-1 and type-2 detectors, respectively.  The sidelobes increase the beam solid angle compared to the best fit Gaussian by 30\%.  The effective beam area relative to the Gaussian is factored into the calibration of the instrument (Sec.~\ref{sec:calibration}).

\begin{figure}[th]\centering
\includegraphics[width=3.375in]{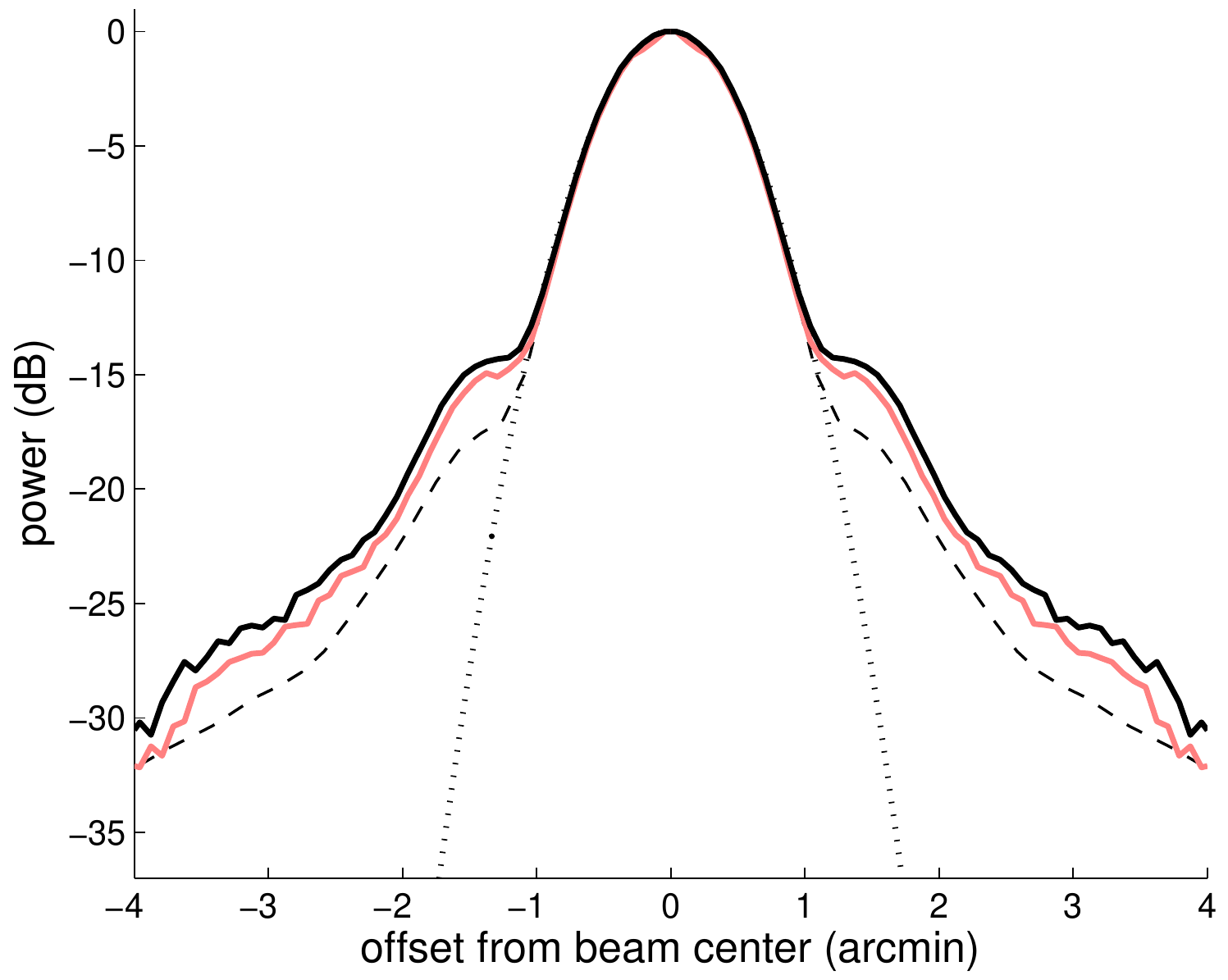}
    \caption[]{Comparison of measured versus predicted average beam pattern of type-1 and type-2 detectors. The individual channel maps from a planet observation are coadded and radially averaged to derive an average beam profile for the type-1 (black solid line) and type-2 (red solid line) detectors.  The average predicted beam, also radially averaged (dashed lines), shows near sidelobes due to truncation at the Lyot stop.   The measured beam central lobes are well fit by a 59\asec\ Gaussian (dotted line).  The near sidelobe level  \db{-14} (\db{-15}) of the type-1 (type-2) profile is consistent with optical crosstalk of 2.5\% ($<1\%$) between adjacent bolometers.}
    \label{fig:theoreticalbeam}
\end{figure}

We compare the beam profiles of both detector types to the theoretical prediction in Fig.~\ref{fig:theoreticalbeam}.  Theoretical beam shapes for different positions in the array are calculated using \textsc{zemax}.  The theoretical profile in Fig.~\ref{fig:theoreticalbeam} is calculated from a weighted composite of the predicted beams from several positions in the array.  The predicted beam has sidelobes due to truncation at the Lyot stop.  The measured beams show increased near-sidelobe power over that predicted which is consistent with optical crosstalk between adjacent bolometers.  The sidelobe level of the type-1 (type-2) profile indicates $2.5\%$ ($<1\%$) crosstalk between adjacent bolometers.  This level of crosstalk and the factor 2.5 difference in crosstalk is consistent with cavity simulations (Fig.~\ref{fig:hfss}).

\subsection{Band}
\label{sec:perf.band}

\begin{figure}[th]\centering
\includegraphics[width=3.375in]{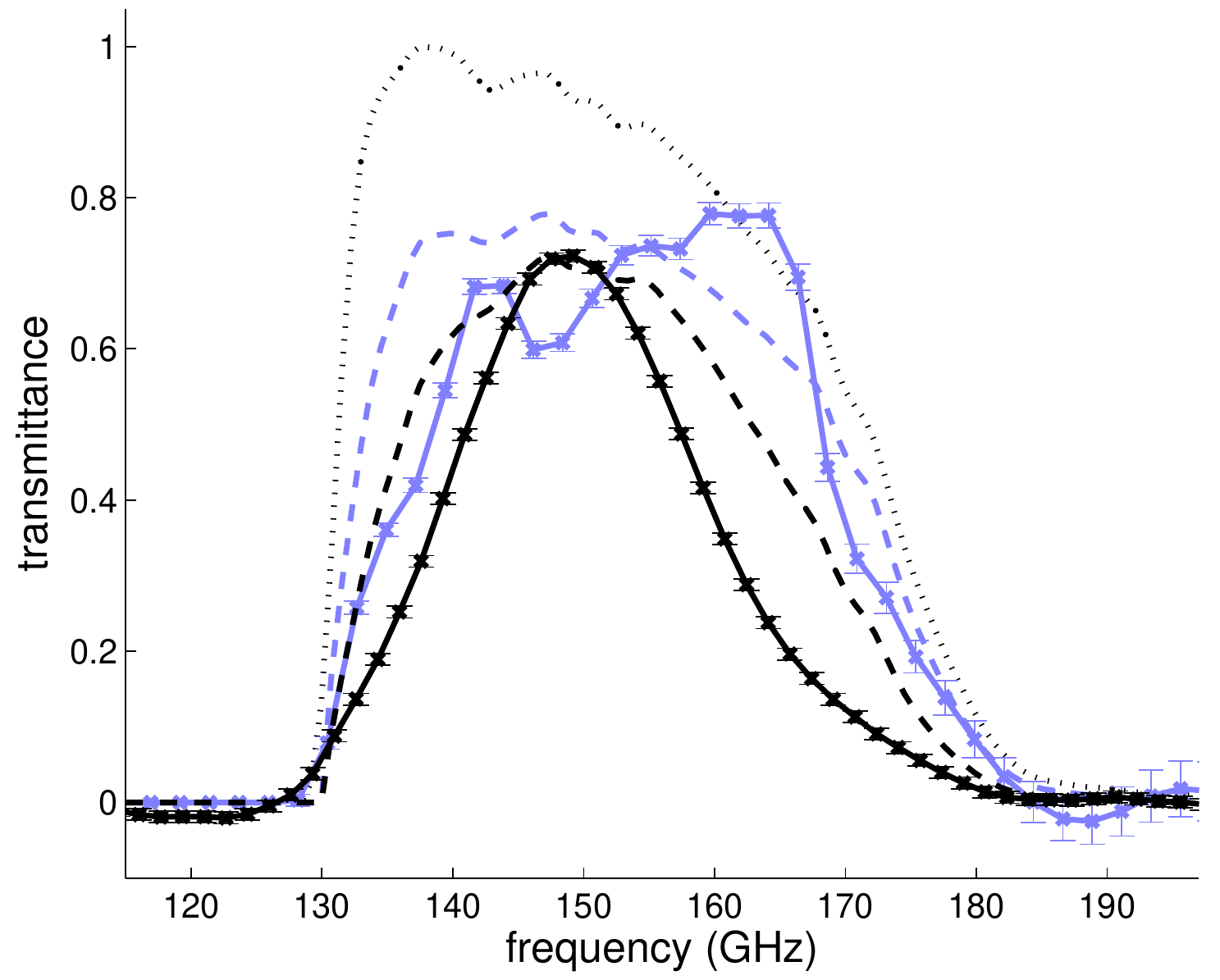}
    \caption[]{Bandpass for a perfect cavity (dotted line) and that of
 type-1 (black) and type-2 (blue) cavity geometries.  The bandpass of the perfect cavity is calculated from the mesh filter$\times$waveguide transmittance.  The type-1~and~2 detectors have curves for the FTS measured bandpass (solid lines with uncertainties) and the calculated mesh filter$\times$waveguide$\times$cavity transmittance (dashed lines). Estimated statistical measurement uncertainties are shown; these are not uniform for the measured type-2 spectrum due to post-measurement corrections for differences between the filters in the test setup and those in the field. In addition, there are systematic uncertainties due to the optical coupling geometry between the detectors and the FTS which may cause ripples in spectrum.  The FTS measured results are normalized to the peak calculated transmittance of the corresponding detector type.  The absorbed power of each  configuration is proportional to the area under the curve.  The measured type-1 bandpass is much narrower than expected, while the measured and calculated bandpass of the type-2 configuration are more similar.}
     \label{fig:fts}
\end{figure}

The expected bandpass for the system is calculated from the
transmittance spectrum of the metal-mesh LP filters and the HFSS
simulation of the waveguide, and cavity geometry.  The spectral
bandpass of sample channels on type-1 and type-2 sub-arrays were
measured using a Fourier transform spectrometer (FTS).  The calculated
and measured bandpass functions for both detector types are shown in
Fig.~\ref{fig:fts}.

Each spectrum is characterized by an effective bandwidth, center frequency, and peak value.  The effective bandwidth is given by
\begin{equation}
\veff=\int{F(\nu)d\nu},
\label{eq:veff}
\end{equation}
where $F(\nu)$ is the bandpass function normalized to a peak value of unity.  The center frequency is
\begin{equation*}
\nu_c =\frac{\int{F(\nu) \nu~ d\nu}}{\int{F(\nu)~d\nu}}.
\label{eq:vcen}
\end{equation*}

The expected loss in efficiency due to the type-1 (type-2) absorbing
cavity is estimated by comparing the calculated spectrum for the mesh
filters, waveguide, and cavity to that assuming a
perfect absorbing cavity (i.e, mesh filters and waveguide only).  We
then define the cavity efficiency, $\eta_{cav}$, as the ratio of the
peak transmittances of these two calculated spectra. The cavity
efficiency is 0.72 and 0.78 for the type-1 and type-2 detectors,
respectively.

\begin{table}[th]
\begin{center}
\begin{ruledtabular}
\begin{tabular}{lccc} 
                                    & \multicolumn{2}{c}{Calculated} & Measured\\ 
                     Wafer geometry & $\eta_{cav}$ & \veff\ (GHz) & \veff\ (GHz) \\ 
\colrule 
                               Perfect cavity &          1 &       37.6 & 	  \ldots \\ 
                                       Type-1 &       0.72 &       32.3 &       24.5 \\ 
                                       Type-2 &       0.78 &       37.5 &       33.5 \\ 
\end{tabular} 
\end{ruledtabular}

\end{center}
\caption{Cavity efficiency and effective bandwidth calculated for predicted and measured band pass spectra for the two detector types.}
\label{tbl:nueff}
\end{table}

The cavity efficiency as well as calculated and measured effective bandwidths are given in Table~\ref{tbl:nueff}.  Note that the FTS measurements are in relative units, so there is no direct measurement of $\eta_{cav}$.  The design bandwidth is 37.6~GHz.  Taking into account the frequency dependence of simulated cavity absorption, the expected effective bandwidth of the type-1 detectors is 32.3~GHz, while their measured \veff\ is 24.5~GHz.   The anomalously narrow band adversely affects the overall sensitivity of the type-1 detectors (Sec.~\ref{sec:perf.sens}).  In contrast, the measured bandwidth of the type-2 detectors, 33.5~GHz, is much closer to the corresponding calculated bandwidth, 37.5~GHz.   While we do not fully understand the below-expected performance of the type-1 sub-arrays, the type-2 detector performance approaches that expected from simulations.

Infrared leaks in the LP mesh filters can contribute significantly to the detector optical loading.  We place upper limits on the high frequency out-of-band response of the instrument in the laboratory with measurements using a series of high-pass thick grille filters with cutoffs at 146~GHz, 206~GHz, and 305~GHz.  The signal from a hot-chopped source sets an upper limit of 3\sci{-4} for the ratio of out-of-band to in-band response to an Rayleigh-Jeans (RJ) temperature source.

\subsection{Calibration}
\label{sec:calibration}

The detector timestream data are recorded in counts from the ADC which are proportional to the bolometer current.  We make a daily raster scan of an astronomical source of known brightness to measure the conversion from ADC units to astronomical source flux.

Mars is our most frequently used calibration source.  The brightness temperature of Mars is calculated from the Rudy model\cite{rudy1987,*muhleman1991} scaled by a factor 1.052.   The scaling factor is derived by comparing the Rudy model results at 93~GHz with Wilkinson Microwave Anisotropy Probe (WMAP) measurements of Mars at 93~GHz (additional details in Ref.~\onlinecite{halverson2009}).  When Mars is not available, we observe the RCW38 HII region, which we have calibrated at 150~GHz with Mars.

\subsubsection{Calibration Procedure}
\label{sec:cal.proc}

We calculate a factor to convert ADC counts to changes in RJ temperature, $\Delta T_\rj$, on each detector for each night of observation.  The equivalent RJ temperature change on the sky seen by detector $n$ is
\begin{equation}
\Delta T_{\rj,n} = \frac{\Delta P_{S,n}}{2 k_B \veff},
\label{eq:trj}
\end{equation}
where $k_B$ is the Boltzmann constant, \veff\ is given by Eq.~\eqref{eq:veff}, and $\Delta P_{S,n}$ is the expected optical power difference between the calibration source $S$ and the CMB for a single-moded beam.  The individual calibration scan maps from each detector are corrected for the finite source size and fit to an elliptical Gaussian beam as described in Sec.~\ref{sec:beam.gauss}.  Then, the calibration factor for each detector is
\begin{equation}
a_{T_\rj,n} = \frac{\Delta T_{\rj,n}}{A_n},
\label{eq:calf}
\end{equation}
where $A_n$ is the amplitude of the Gaussian fit to the uncalibrated source map.

Determining the expected power from the source, $\Delta P_{S,n}$, requires a measurement of the beam pattern.  As detailed above (Sec.~\ref{sec:beam.prof}), we cannot accurately measure the beam solid angle for each beam.  We, therefore, use the mean beam profile (Fig.~\ref{fig:theoreticalbeam}) and assume a fixed ratio of the true beam solid angle to the best-fit Gaussian beam,
\begin{equation*}
\Omega=\int_{2\pi} P_n~ d\ohm \approx \rho \int_{2\pi} G_n~ d\ohm.
\end{equation*}
Here $P_n$ and $G_n$ are the normalized beam and elliptical best fit Gaussian power patterns, respectively, and $\rho$ is the scaling factor between the two, for example, $\rho = 1.32$ for type-1 detectors. The best-fit Gaussian beam solid angle for each channel is increased by this factor.  With this approximation, the beam dilution factor for a compact source, such as Mars, is calculated as
\begin{equation}
\eta_{S,n} =\frac{\int_S G_n~ d\ohm}{\rho \int_{2\pi} G_n~ d\ohm},
\label{eq:beamrhoapprox}
\end{equation}
where the integral in the numerator is taken over the solid angle of the calibration source.

With the above form for $\eta_{S,n}$, $\Delta P_{S,n}$ is
\begin{equation}
\Delta P_{S,n} = \eta_{S,n}c^2 \int \frac{B(\nu,T_S)-B(\nu,T_{CMB})}{\nu^2} F(\nu) d\nu,
\label{eq:calib}
\end{equation}
where $B(\nu,T)$ is the Plank blackbody brightness spectrum, $T_S$ is the source temperature, and $T_{CMB}$ is 2.73~K.  For the beam measurement and calibration of August detailed here, Mars was used with $T_S = 207 \pm 3$~K

The calibration factor is calculated from equations \eqref{eq:trj},~\eqref{eq:calf},~\eqref{eq:beamrhoapprox},~and~\eqref{eq:calib}.
Finally, we apply an elevation dependent atmospheric opacity correction to each scan.  The correction has the form
\begin{equation*}
\chi = e^{\tau \csc\left(\varepsilon_{scan}\right)}e^{-\tau \csc\left(\varepsilon_{cal}\right)},
\end{equation*}
where $\tau$ is the optical depth at zenith, $\varepsilon_{scan}$ is the median elevation of the scan to be calibrated and $\varepsilon_{cal}$ is the elevation of the calibration scan.  Section~\ref{sec:skydips} describes the measurement of $\tau$.  The correction factor is generally $<3\%$.

\subsubsection{Atmospheric Opacity}
\label{sec:skydips}

Atmospheric emission is measured with skydips in which the telescope is scanned from zenith to low elevation on each night of observation.
Modeling the atmosphere as a horizontal layer of emission produces a sky signal which will vary with observation elevation ($\varepsilon$) as
\begin{equation}
\Delta T_\rj(\varepsilon) = T_{atm}\left(e^{-\tau} - e^{-\tau \csc(\varepsilon)} \right),
\label{eq:taufit}
\end{equation}
where $T_{atm}$ is the temperature of the atmosphere and $\tau$ is the optical depth at zenith.

\begin{figure}[th]\centering
\includegraphics[width=3.375in]{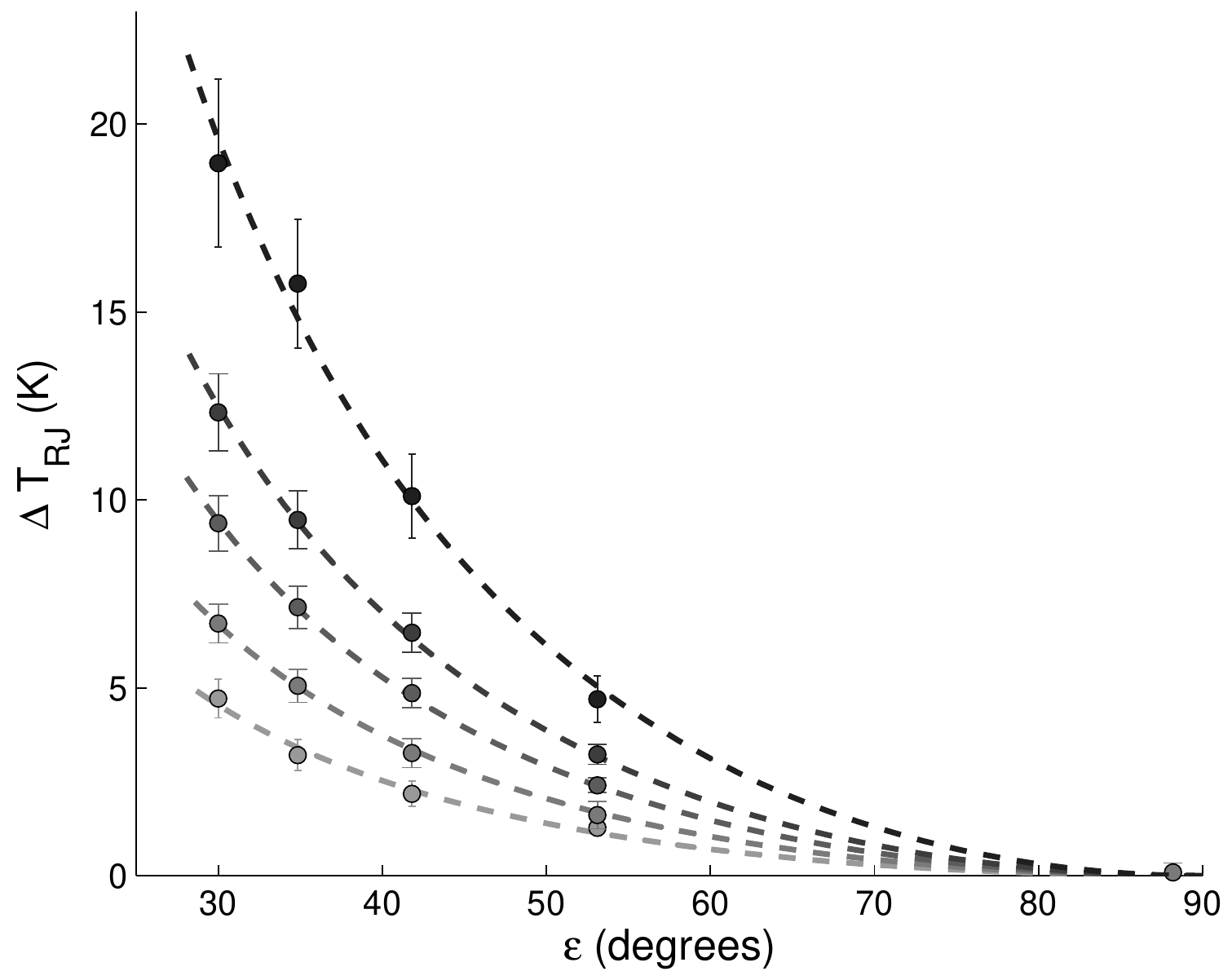}
    \caption[Skydip scans of varying opacity]{Skydip scans of varying opacity.  We fit each scan according to Eq.~\eqref{eq:taufit} to find $\tau$, assuming an atmospheric temperature of 273~K. The curves shown here have, from top to bottom, $\tau =$ 0.072, 0.046, 0.035, 0.025, 0.017.}
    \label{fig:opac}
\end{figure}

In fitting Eq.~\eqref{eq:taufit}, $\tau$ and $T_{atm}$ are largely degenerate.  We lack reliable atmospheric temperature data, so we assume $T_{atm}=273$~K.  Figure~\ref{fig:opac} shows five different skydips for a range of optical depths taken in August 2007 along with model fits.   In the fits, each channel is fit individually, but the plots show the median response at each elevation with a curve set by the median $\tau$. The skydips are well fit down to an elevation of 30\dg, indicating that the detector response remains linear over this range of opacities.

In the August 2007 observing period, the optical depth at zenith ranged between 0.017 and 0.072, with a median of 0.032.  Using the APEX atmospheric transmittance calculator and the \apex\ spectral bandpass, we calculate a PWV of 1.4~mm for the median optical depth.  We extrapolate the results to zero airmass and determine the background load contribution of the atmosphere.  There is an atmospheric load of $\sim1.3$~pW on the detectors at 55\dg\ elevation when the optical depth is 0.032.

\subsubsection{Calibration Stability}
\label{sec:cal.stab}

The calibration uncertainty is determined by comparing the calibration from separate days.  For seven successive days, we fit the coadded Mars map 
to a Gaussian.   Among the seven maps, there is a $\pm 1.9\%$ variation in fit amplitude after correcting for differences due to Mars temperature and atmospheric opacity.  This variation is due to small changes in bolometer bias parameters from night to night.

We typically focus the secondary mirror and perform only one large array calibration raster for each night of observation.  Focus drifts change the relative calibration of detectors but have little effect on the average array calibration.  Defocusing of beams in one region of the array is largely compensated by focusing the opposite region.

The measured array responsivity can change during the night due to changes in the temperature of the detector cold stage or atmospheric loading on the detectors.  Short scans of calibration sources are used to monitor gain stability over the course of the night. In a sample series of Mars observations taken over 4~h, there was a $\pm1.6\%$ brightness variation among scan maps after correcting for elevation changes.  Therefore, while the average calibration on a given night is known to 1.9\%, the average detector gain for an individual scan differs by an additional 1.6\%.

\subsection{Pointing}
\label{sec:pointing}

A pointing model was developed for APEX from optical observations of stars.  The measured pointing error is $\pm 2\asec$ and tracking accuracy is $\pm 0.6\asec$ under stable atmospheric conditions.\cite{gusten2006}  However, we frequently observe through sunset and sunrise during which temperature changes cause slow drifts in the pointing.  Such drifts increase the effective beam size in maps coadded from multiple scans.  During observations, we check the pointing by scanning a planet or quasar every 1--2~h or when changing targets.  The applied corrections have an rms variation of $\sim4\asec$, which we take to be the pointing uncertainty.

Measurements of the positions of the individual beams relative to the center of the focal plane are very stable.  The rms variation for Mars scans, which vary 20\dg\ in elevation, is 1.1\asec.  No correction is necessary.  The total pointing uncertainty for the beams is $\pm4.1\asec$.  The correction to the calibration from this uncertainty is less than $0.5\%$, which is negligible.

\subsection{Instrument Efficiency}
\label{sec:perf.eff}

The change in power measured in the bolometer, $\Delta P_{opt}$, and the expected power from a source, $\Delta P_S$ in Eq.~\eqref{eq:calib}, differ because of losses in the optical elements, loss of thermal energy in the spiderweb, imperfect cavity absorption, and uncertainty in the conversion of thermal to electrical power in the bolometer.
We define an efficiency associated with each of these elements: $\eta_{opt}$, $\eta_{bolo}$, $\eta_{cav}$, and $\eta_{resp}$, such that
\begin{equation*}
\eta_{tot} = \eta_{opt} \; \eta_{bolo} \; \eta_{cav} \; \eta_{resp}.
\end{equation*}
The optical efficiency, $\eta_{opt}$, is calculated from the expected reflection and scattering of optical elements.  The transmittance of individual elements are listed in Table~\ref{tbl:optloads}.  However, 34\% of the throughput of each feed horn is truncated at the Lyot stop. Since this is an element of the optics design rather than a loss due an imperfect optical element, we compensate for this loss in the efficiency calculation.  By this measure, a perfectly lossless optical path which includes a Lyot stop has $\eta_{opt}=1$.  The total transmittance of the optical elements other than the Lyot is $\eta_{opt}=0.62$.

The bolometer absorption efficiency, $\eta_{bolo}$, is discussed in Sec.~\ref{sec:yield}.  Calculation of the cavity efficiency, $\eta_{cav}$, is described in Sec.~\ref{sec:perf.band}.

The responsivity efficiency, $\eta_{resp}$, can deviate from unity due to finite ETF loop gain as shown in Eq.~\eqref{eq:vbresp}.  The bolometers are biased at a high temperature in the transition and are operated with a loop gain of 5--10.  We estimate that the responsivity differs from the infinite loop gain limit by up to 15\% but do not have an accurate measurement of the loop gain for all detectors.  In addition, the measurement of $V_{bias}$ deviates from the actual bias voltage applied to the bolometer by up to 10\% due to phase shifts in the cryostat wiring and residual impedance of the $LC$
resonator. However, these two factors have opposing effects on the power responsivity.  As a rough approximation, we assume that $\eta_{resp}$ is unity.

The cumulative efficiency with which the instrument converts changes in source power to electrical power in the detector, $\eta_{tot}$, is measured by the ratio of $\Delta P_{opt}$ to $\Delta P_S$.   Thus, the cumulative efficiency is given by
\begin{equation}
\eta_{tot} =  \frac{\Delta P_{opt}}{\Delta P_S (1-L_{Lyot})},
\label{eq:etaopt}
\end{equation}
where $L_{Lyot} = 0.34$.  The median measured \apex\ efficiency is $\eta_{tot}=0.31$ for type-1 detectors and 0.36 for type-2. The difference in efficiency is predicted by simulations of the two backshort geometries.

\subsection{Optical Loading}
\label{sec:perf.loading}

We measure the total background optical load on the bolometers by comparing the detector current as a function of bias measured during observation with those of dark detectors.  Dark voltage response curves were measured in the laboratory before deployment.  The median total optical load on the type-1 bolometers during the August 2007 observations is 6.1~pW.  This corresponds to a loading temperature $T_\rj= 44$~K.

The estimated contribution to the optical load for each optical element is listed in Table~\ref{tbl:optloads}.  The atmospheric load is the measured value described in Sec.~\ref{sec:skydips}.  The total estimated load is 6.5~pW, and estimated efficiency is 0.36 (~$0.24/(1-L_{Lyot})$~).  Both are greater than the measured values.  The most likely source of error is $\eta_{resp}$.  An $\eta_{resp}$ of 0.9 would result in a calculated load and efficiency consistent with the measured values.

\begin{table*}[th]
\begin{center}
\begin{ruledtabular} 
\begin{tabular}{lcccccc} 
                  Element &  $T_e$ (K) & $\epsilon$ &      $L_s$ &  $T_s$ (K) & $\eta_{e}$ & $P_{opt}$ (pW) \\ 
\colrule 
                      CMB &       2.73 &          1 &          0 &        2.7 &       1.00 &      0.089 \\ 
               atmosphere &        273 &       0.03 &          0 &        270 &       0.97 &        1.3 \\ 
           primary mirror &        273 &       0.02 &      0.011 &         10 &       0.97 &       0.93 \\ 
         secondary mirror &        273 &      0.003 &    6.4e-04 &        140 &       1.00 &       0.14 \\ 
                     M1SZ &        300 &      0.003 &      0.011 &        300 &       0.99 &       0.73 \\ 
                     M2SZ &        300 &      0.003 &      0.011 &        300 &       0.99 &       0.74 \\ 
                     M3SZ &        300 &      0.003 &      0.011 &        300 &       0.99 &       0.75 \\ 
                   window &        300 &          0 &     0.0047 &        300 &       1.00 &       0.25 \\ 
        12 THz LP blocker &         70 &          0 &       0.01 &         64 &       0.99 &       0.11 \\ 
        3.0 THz LP filter &         70 &          0 &       0.05 &         51 &       0.95 &       0.46 \\ 
        2.4 THz LP filter &         70 &          0 &       0.05 &         38 &       0.95 &       0.35 \\ 
        255 GHz LP filter &          6 &          0 &       0.05 &          6 &       0.95 &      0.033 \\ 
                Lyot stop &          6 &          0 &       0.34 &          6 &       0.66 &       0.34 \\ 
               field lens &          6 &       0.02 &       0.04 &          6 &       0.94 &      0.063 \\ 
             imaging lens &          6 &       0.07 &       0.04 &          6 &       0.89 &       0.13 \\ 
        177 GHz LP filter &       0.28 &          0 &       0.05 &       0.28 &       0.95 &    1.6e-12 \\ 
                     horn &       0.28 &          0 &          0 &       0.28 &       1.00 &          0 \\ 
            type-1 cavity &       0.28 &         \ldots &         \ldots &         \ldots &       0.72 ($\eta_{cav}$) & \ldots \\ 
                bolometer &       0.28 &         \ldots &         \ldots &         \ldots &       0.84 ($\eta_{bolo}$) & \ldots \\ 
             responsivity &       0.28 &         \ldots &         \ldots &         \ldots &          1 ($\eta_{resp}$) & \ldots \\ 
\\ 
                    Total &            &            &            &            &       0.24 &       6.40 \\ 
\end{tabular}
\end{ruledtabular} 

\end{center}
\caption{Optical loading and loss for individual optical elements.  $T_{e}$ is the physical temperature of each element, $\epsilon$ is the emissivity, $L_{s}$ is the spillover/scattering loss of each element, $T_{s}$ is the temperature of the spillover/scattered radiation absorber, $\eta_e$ is the estimated transmittance/efficiency of each element, and  $P_{opt}$ is the contribution to total background optical power absorbed by the bolometer.  Results are tabulated for type-1 bolometers.  The total load corresponds to an RJ loading temperature of 44~K.}
\label{tbl:optloads}
\end{table*}

Atmospheric loading is the largest single component of the background optical load.  The sky and telescope contribute 40\% of the total loading.  Significant contributions also come from the reimaging mirrors and the 60~K thermal filters.  The three mirrors inside the cabin have high quality surfaces, but they scatter light inside the cabin to 300~K.  Together, the mirrors contribute roughly 35\% of the total optical load.

\subsection{Noise and Sensitivity}
\label{sec:perf.sens}

\subsubsection{Instrument Noise}

\begin{table*}[th]\centering
\begin{center}
\begin{ruledtabular}
\begin{tabular}{lccc} 
             Noise Source &   Equation & \dcoltext{Type-1 NEP (aW/\rthz)} & \dcoltext{Type-2 NEP (aW/\rthz)} \\ 
\colrule 
\em{Readout} \\ 
         Warm electronics & $3.7 \times \pi/2\,[\mbox{pA}/\rthz] \times V_{bias}/\sqrt{2}$ &         25 &         14 \\ 
                    SQUID & $2.5\times \pi/2\,[\mbox{pA}/\rthz] \times V_{bias}/\sqrt{2}$ &         17 &        9.2 \\ 
Bias resistor Johnson noise & $1.4 \times \sqrt{2} \,[\mbox{pA}/\rthz] \times V_{bias}/\sqrt{2}$ &        8.5 &        7.9 \\ 
\\ 
\em{Bolometers} \\ 
    Thermal carrier noise &                      $\sqrt{\gamma 4 k_B T_c^2 G}$ &         36 &         33 \\ 
            Johnson noise &                       $\sqrt{4 k_B T_c P_{bias} }$ &         24 &         19 \\ 
        Photon shot noise &                          $\sqrt{2 h \nu P_{opt} }$ &         35 &         50 \\ 
    Photon bunching noise &                    $\sqrt{\xi {P_{opt}^2}/ \veff}$ &         39 &         66 \\ 
\\ 
          Total ($\xi=0$) &                                                    &         64 &         65 \\ 
          Total ($\xi=1$) &                                                    &         75 &         92 \\ 
\end{tabular} 
\end{ruledtabular}

\end{center}
\caption{Summary of contributions to NEP.  Median values of the array are used to calculate the expected noise contributions from  readout, detectors and optical loading.  The photon bunching noise is listed as an upper limit corresponding to $\xi=1$.}
\label{tbl:noise}
\end{table*}

Table~\ref{tbl:noise} summarizes the expected median noise contributions from readout, detectors, and optical load for a single detector.  Values used to calculate the noise terms are the median values of the array for type-1 (type-2) detectors: $V_{bias}=6.0$ (3.3) ~$\mu$V, $P_{bias}=22$ (11)~pW, and $P_{opt}=6.1$ (12)~pW.

The SQUID and warm readout electronics (readout amplifier, feedback resistor, and nulling resistor) produce broadband current noise which appears incoherently in both sidebands of the carrier.  When the bolometer signal is recovered, the sidebands are summed and the noise from both contribute.  In addition, these signals do not pass through the $LC$ filters so, when demodulated with the square-wave mixer, there are additional contributions from noise at odd harmonics of the carrier band.  The total noise is a factor $\pi/2$ larger than the nominal noise level of these components.  We refer these noise current terms to a noise equivalent power (NEP) via the TES responsivity, Eq.~\eqref{eq:vbresp}.  Note that the ratio of signal to readout noise for this system, which uses alternating voltage bias and a square-wave demodulator, is within $10\%$ of that for a constant voltage bias detector. This ratio would be unity if the demodulator in this system used a sine-wave mixer, as is the case for the digital multiplexer.\cite{dobbs2008}

The bolometers have a noise contribution from the fluctuation of thermal carriers in the bolometer thermal link.  The thermal carrier noise has the form given in Table~\ref{tbl:noise}.  The factor $\gamma$ accounts for a temperature gradient along the link.\cite{mather1982}  For a normal metal link at the detector operating temperatures, $\gamma=0.58$.   The Johnson noise from the TES and bias resistor also appear incoherently in both sidebands of the carrier.  These signals are filtered by the $LC$ filters so there are no additional harmonic contributions. The total noise is a factor $\sqrt{2}$ larger than the nominal Johnson noise of a resistive element.  When operated at high loop gains, ETF suppresses noise current though the bolometer, including Johnson noise.  Table~\ref{tbl:noise} includes the Johnson noise at the full value, which should be taken as a worst case.

We include two terms for the photon noise from background loading: shot noise and ``bunching'' noise.\cite{lamarre1986}  The bunching term takes into account the boson nature of photons and has an uncertain magnitude which we parameterize as $\xi \in [0,1]$.\cite{runyan.thesis}  The expected total detector white noise is in the range 64--75 and 65--92~aW/\rthz\ for type-1 and type-2 detectors, respectively.

In the type-1 detectors, the total readout noise, bolometer thermal noise, and photon noise are all comparable.  The narrow bandwidth and 0.72 cavity efficiency of the type-1 detector reduce the incident optical power.  As a result, the thermal conductance, chosen based on the expected load, is unnecessarily high.  The high $G$ results in a high thermal carrier noise and increases the $V_{bias}$ required to maintain the detectors in the superconducting transition.  Thus, all the non-optical noise terms are elevated.  The electrical bias power required to hold the bolometers in the superconducting transition is 3.5 times the absorbed optical power.

Type-2 detectors move toward background limited performance with increased bandwidth and optimized bolometer properties.  The wider bandwidth results in larger $P_{opt}$ and increased photon noise.  The lower $G$ decreases the required $V_{bias}$ and $P_{elec}$ which decreases all non-photon noise contributions (see Table~\ref{tbl:boloprop}).  The lower $R_n$  decreases the required $V_{bias}$ which results in lower readout noise.  In the type-2 detectors, applied bias power and absorbed optical power are roughly equal.

\subsubsection{Detector Sensitivity}

The achieved bolometer NEP is the white noise level in the signal band after atmospheric noise removal (Sec.~\ref{sec:obs}).  The signal band is the white noise region between $1/f$ noise and the bolometer response roll off.  After atmospheric noise removal, the bolometer timestreams have $1/f$ noise with a knee at 1--2~Hz which is dominated by the detector readout.  We take 3~Hz to be the minimum frequency which is safely above the $1/f$.  The bolometer time constant (Table~\ref{tbl:boloprop}) filters signal frequencies $f > 1/(2\pi \tau_{opt})$.  The resultant signal band is 3--14~Hz and 3--8~Hz for the type-1 and type-2 detectors, respectively.  When observing, the telescope scan speed is selected so the modulated sky signal appears in the signal band (Sec.~\ref{sec:scan}).

\begin{figure}[th]\centering
\includegraphics[width=3.375in]{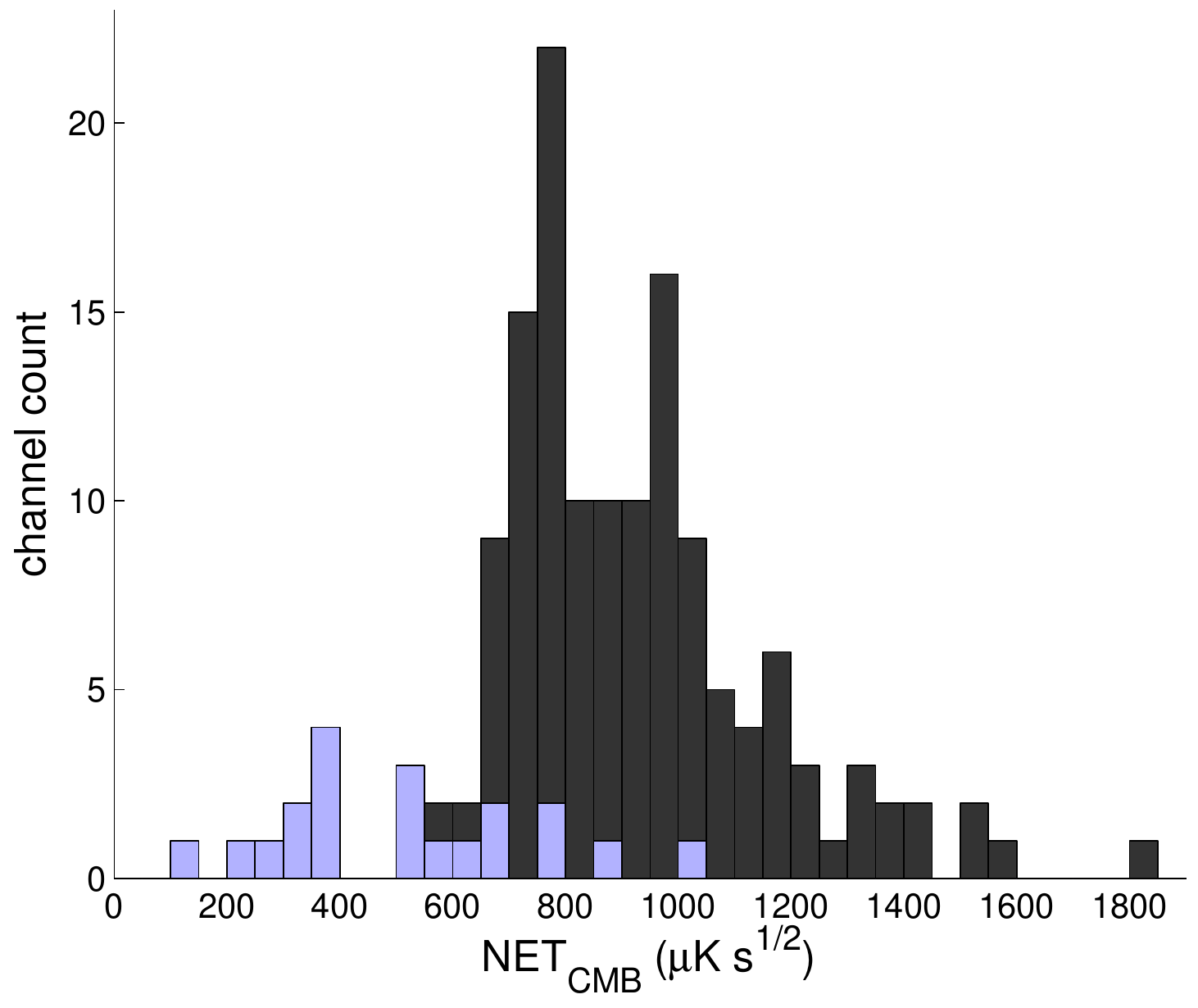}
    \caption[]{Histogram of individual channel sensitivities in the array.  Sensitivities are measured in the signal band of each detector type after analysis pipeline atmospheric removal and data cuts. NET values are plotted for the 135 type-1 channels (dark gray) which remain after all data cuts during an observation of the Bullet cluster.  Also included are 20 channels from a type-2 sub-array (blue) measured in a separate observation.  The type-2 histogram does not obscure any type-1 values.}
    \label{fig:hist}
\end{figure}

Table~\ref{tbl:net} lists the median sensitivity of the bolometers in flux and temperature units, while  Fig.~\ref{fig:hist} shows a histogram of the pixel sensitivities of the type-1 array and a type-2 sub-array.  The values are calculated from bolometer channels which remain after analysis pipeline data cuts.

\begin{table}[th]\centering
\begin{ruledtabular}
\begin{tabular}{lcc} 
              Sensitivity &     Type-1 &     Type-2 \\ 
\colrule 
                      NEP (aW/\rthz)&         97 &         87 \\ 
                     NEFD (mJy$\sqrt{s}$)&         42 &         24 \\ 
                      NET ($\mu\text{K}_\rj \sqrt{s}$)&        500 &        300 \\ 
                      NET (\ukcmbs)&        890 &        530 \\ 
                    NE$y$ (\sci{-4}\,$\sqrt{s}$) &        3.5 &        2.2 \\ 
\end{tabular}
\end{ruledtabular}

\caption{Observed median noise and sensitivity per channel.  NEP is the measured noise equivalent power in the detector.  NEFD is the noise equivalent flux density, measuring the sky-signal sensitivity. The NET values are the noise equivalent temperature referred to a source at the RJ limit and the CMB temperature.  NE$y$ is the noise equivalent $y$, the dimensionless Comptonization parameter. }
\label{tbl:net}
\end{table}

The type-1 detectors have a median NEP of 97~aW/\rthz, median NET of 890~\ukcmbs\ and median NE$y$ of 3.5\sci{-4}~$\sqrt{s}$.  Here, $y$ is the dimensionless Comptonization parameter, proportional to the cluster optical depth.  The type-1 NEP is well above the maximum expected noise of 75~aW/\rthz.  Approximately one third of the channels have an NEP in the expected range.  These are the channels with lower bias frequencies.  As described in Sec.~\ref{sec:readout}, there are frequency dependent phase shifts between the carrier and output demodulator in the readout system.  The phase shifts result in the sensitivity being dependent on bias frequency which is not accounted for in Table~\ref{tbl:noise} and which elevates the detector noise above the expected level for high bias frequencies. The lower optical load on the type-1 detectors enhances the frequency dependent excess noise because the non-optical noise terms are more significant.  This frequency dependent effect increases the variance in detector sensitivity of Fig.~\ref{fig:hist}.  The long tail of high NET detectors is due to a subset of detectors with low optical efficiency.

The NEP of the type-2 detectors, 87~aW/\rthz, is within the expected range.  The type-2 detectors also show elevated noise with increasing bias frequency, but the effect is smaller, and the median NEP falls
within the rage of expected values.  The type-2 detectors have a median NET of 530~\ukcmbs\ and median NE$y$ of 2.2\sci{-4}~$\sqrt{s}$.  Overall, the type-2 sub-arrays have median detector sensitivities a factor 1.7 better than the type-1 design.  This improvement is consistent with the wider measured bandwidth and higher cavity efficiency of these detectors.

\section{Observations and Analysis}
\label{sec:obs}

We operate the receiver on a 24~h cycle, in line with telescope sun avoidance periods and the sorption fridge hold time.  Each night, we check the secondary mirror focus position, perform 1--2 skydips to characterize the atmosphere (Sec.~\ref{sec:skydips}), and make periodic pointing checks (Sec.~\ref{sec:pointing}).  Approximately 10\% of our observation time is used for flux calibration, including at least one full raster of a known source for calibration and beam measurement and periodic scans of secondary calibrators to monitor gain stability through the night (Sec.~\ref{sec:cal.stab}).  With these tasks, target switching, and telescope down time, roughly 12~h each night are spent on science targets.

\subsection{Scan Strategy}
\label{sec:scan}

A key goal of \apex\ is the development of a scaling relationship between cluster mass and SZE flux.  To this end, we have focused on targeted observations of known clusters, both relaxed and evolving systems, over a range of redshifts.

\begin{figure}[thbp]\centering
\includegraphics[width=3.375in]{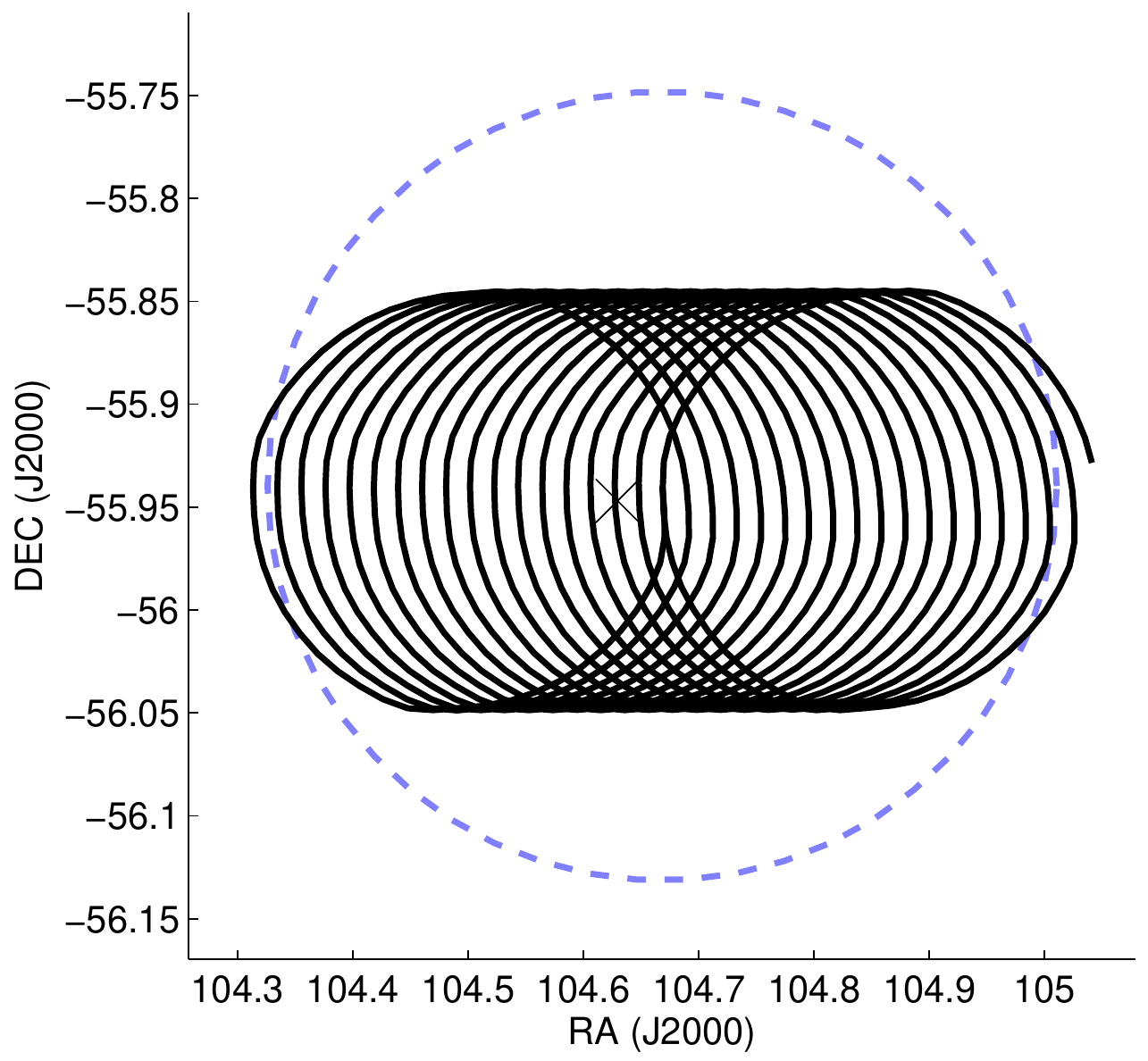}
    \caption[\figtitle]{The circular drift scan pattern used for mapping individual clusters.  The telescope scans in 12\amin\ diameter circles with a 5~s period (black line).  The circle center is stationary in AZ/EL, but tracks in RA/DEC.  After performing twenty circles, the telescope is reset for a subsequent circular drift scan.  Also shown are the FOV of the bolometer array at the initial pointing (blue dashed line) and the source position ($\times$).}
    \label{fig:scan}
\end{figure}

We use circular drift scans to efficiently map a single cluster target within the wide \apex\ FOV.  In the circular drift scan, the telescope makes circular scans which are stationary in azimuth and elevation.   The source drifts across the field with the Earth's rotation.  After the source drifts across the array, the telescope is reset for the new source position.  Figure~\ref{fig:scan} illustrates a typical scan of twenty 12\amin\ diameter circles with a period of 5~s.  This circular scan mode offers a few advantages.  The scan has a continuous, low acceleration, therefore, no data are discarded due to high acceleration turnarounds.  While low acceleration, the scan speed is fast enough ($\sim8$~arcsec/s) that the signal from a typical cluster appears well above the $1/f$ knee of readout and atmospheric noise (Fig.~\ref{fig:psds}).  In addition, the scan is fixed in AZ/EL coordinates allowing us to search for ground signal or other contaminants by summing data from multiple circles.

\subsection{Cluster Map Making}
\label{sec.map}

After completing each scan, the data are recorded as 280 readout channel timestreams sampled at 100~Hz with telescope pointing data interpolated at the same timesteps.  The data reduction pipeline used for generating cluster maps from raw timestreams generally follows a simple sequence: cuts to remove low quality data and optically dead channels, filtering to remove atmospheric noise, cuts to remove data which have residual noise after filtering, and binning filtered data into maps.  In addition to channel cuts (Sec.~\ref{sec:perf.detect}), timestream data are rejected due to step glitches from electrical interference, high telescope acceleration at the start and end of scans, and noisy segments in otherwise good channels.  After all cuts, roughly 40\% of the recorded 280 channel data contributes to the map.

\begin{figure}[thbp]\centering
\includegraphics[width=3.375in]{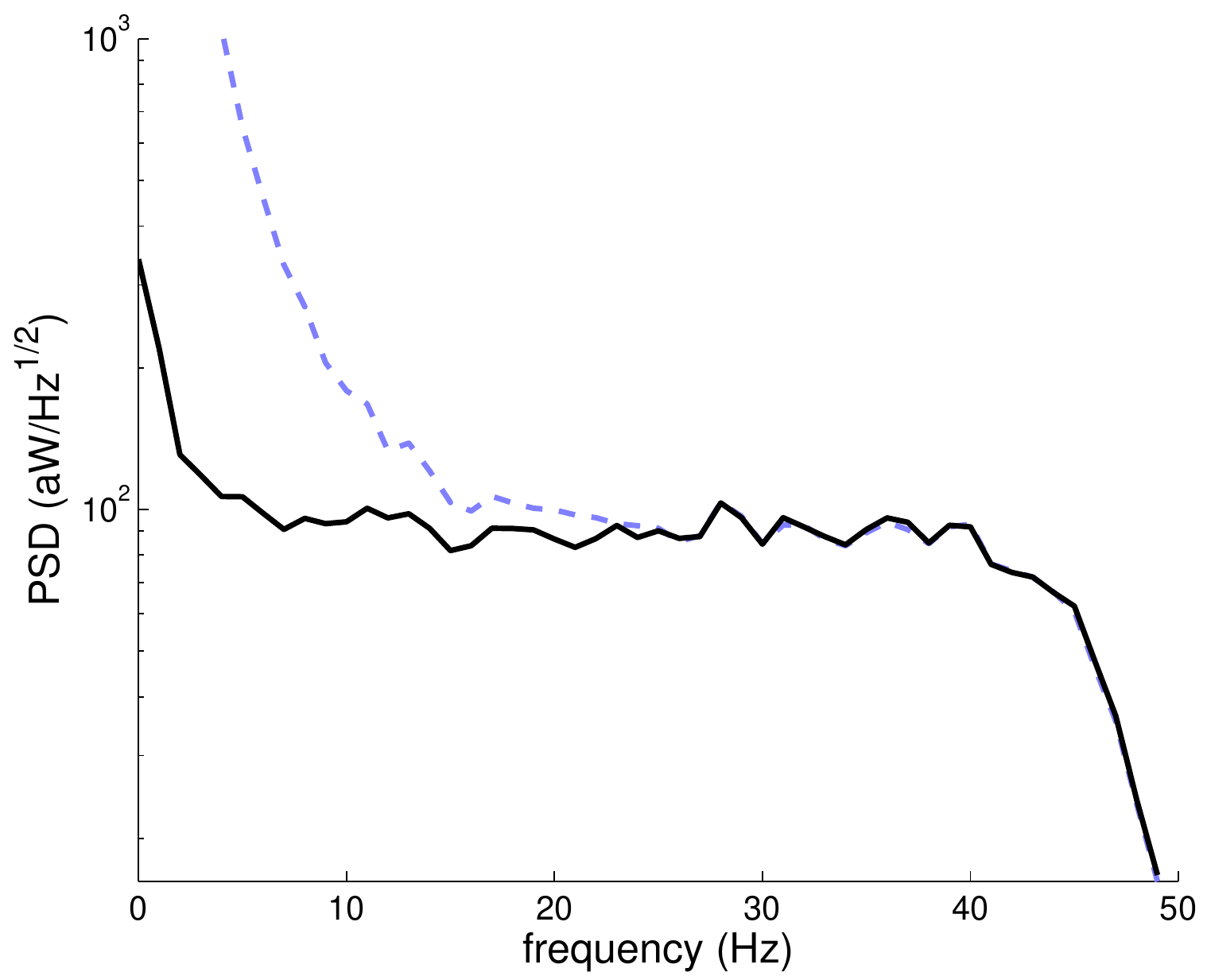}
    \caption[]{Measured power spectral density (PSD) for a type-1 bolometer channel during a single scan made in weather typical of the August 2007 observing run.  The dashed line shows the noise spectrum as measured.  Its $1/f$ knee is 10~Hz.  The solid line shows the same data after atmospheric noise removal by the analysis pipeline.   The $1/f$ knee of the data after atmospheric removal is at 1.4~Hz, below the signal band.}
    \label{fig:psds}
\end{figure}

Atmospheric noise mitigation is based on a low-order time-domain polynomial, which removes slow drifts in the timestreams, and low-order spatial polynomial, which removes signals correlated among bolometer channels according to their relative position in the array.  Figure~\ref{fig:psds} shows noise spectra for a single bolometer during a scan of the Bullet cluster before and after atmospheric noise removal.   After filtering  the large scale correlated noise from the detector time ordered data, the $1/f$ knee of the remaining noise is lowered to $\sim$1~Hz.  Since we can recover the predicted white noise level in the low bias frequency channels, we are confident that we have successfully removed the atmospheric noise in the signal band.

After atmospheric noise removal, the timestream of each channel is binned in a map with the channel AZ/EL offset and flux calibration applied.  These are coadded together with inverse variance weighting to make a map for each scan.  Finally, the individual scan maps are coadded with inverse variance weighting to make the final map of the cluster.

\begin{figure}[thbp]\centering
\includegraphics[width=3.375in]{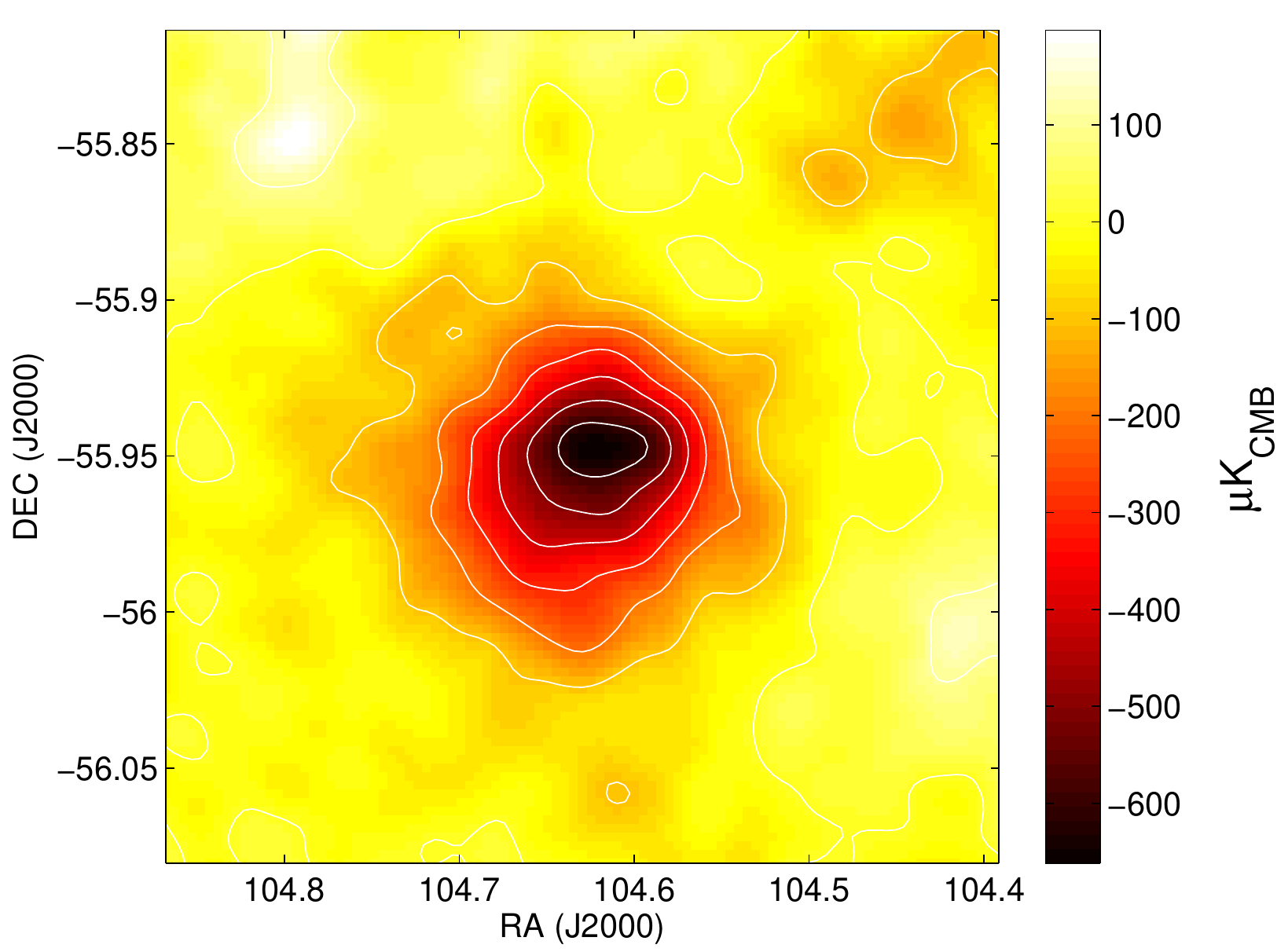}
    \caption[]{SZE map of the Bullet cluster based on 6.5~hr of observation.  Contours are spaces at $100~\mu$K. The map has 85\asec\ resolution with 55~$\mu K_{rms}$ noise.}
    \label{fig:bullet}
\end{figure}

Reference~\onlinecite{halverson2009} details the data reduction and map-making procedures used to create the SZE map of the Bullet cluster shown in Fig.~\ref{fig:bullet}.  The Bullet cluster (1E 0657--56) is a large cluster with total mass $\sim 10^{15}~M_\odot$ and gas temperature $\sim 1.4\sci{8}$~K.  It consists of two merging sub-clusters: the eponymous bullet passing through a larger cluster.  \apex\ observations of the Bullet were made in August 2007.  Because of the large size of the Bullet, the data for this map are processed with the cluster masked in order to preserve signal at the expense of increased noise in the map. After data processing and final smoothing with a 1\amin\ Gaussian kernal, the resulting map has 85\asec\ resolution with 55~$\mu K_{rms}$ noise, resulting in a $23\sigma$ detection in the central 1\amin\ region of the map.

\subsection{Mapping Speed}
\label{sec:mapspped}

The expected mapping speed of the instrument is given by
\begin{equation}
S =  \frac{N_{bolo} \Omega_{beam} \eta_{data}}{\text{NET}^2},
\label{eq:mapspeed}
\end{equation}
$N_{bolo}$ is the number of detectors, $\Omega_{beam}$ is the beam solid angle and $\eta_{data}$ is fraction of collected data which contributes to the map.   Applying the measured detector properties, with $N_{bolo}=280$, $\Omega_{beam}=$1~\sqamin, and $\eta_{obs}=0.4$, we use equation~\eqref{eq:mapspeed} to calculate a nominal mapping speed for the type-1 detector array of 1.4\sci{-4}~deg$^2$/$\ukcmb^2$/hr.

The map of the Bullet cluster is made from the reduction of 234 circular drift scans like that of Fig.~\ref{fig:scan} for a total of 6.5~h of integration time.  The integration time is typical of the \apex\ cluster maps, which require of order 10~h of integration per cluster.  Because of the large FOV of \apex, a significant fraction of the data is taken well away from the region of interest around the cluster.  The equivalent of 3.6~h is spent integrating the central $20\amin \times 20\amin$ area of the map.

The analysis and model fitting of the Bullet cluster in Ref.~\onlinecite{halverson2009} is performed on lower noise map produced without masking the cluster during noise processing.  It has 27~$\mu K_{rms}$ noise over the central $20\amin \times 20\amin$, for a mapping speed of 1.0\sci{-4}~deg$^2$/$\ukcmb^2$/hr or, equivalently, 4.7\sci{-2}~\mapspeed.  There are a few factors which contribute to the discrepancy between the measured and calculated mapping speed. Residual correlated noise likely remains in the map because the extended size of the Bullet requires relatively gentle noise filtering to avoid removing cluster signal.  Additionally, there are several high noise detectors which pass data cuts.  These are included in the integration time, but, when inverse variance weighted, make little contribution to the final map.

Extrapolating from the Bullet results, 30~h of observation on a single target would result in a map with 13~\ukcmb\  noise in the central area.  This integration time is a typical deep integration,  and constitutes a rough practical limit to map sensitivity achievable with \apex.  The sensitivity limit is compatible with our goal of calibrating large SZE surveys, which will map $>1000~\text{deg}^2$ to a depth of 10--20~\ukcmb.

The XLSSU J022145.2-034614 cluster is the lowest mass cluster detected by \apex, with an x-ray temperature of $4.8 \pm 0.6$~keV and estimated mass $M_{500} = 2\sci{14} h^{-1} M_\odot$.\cite{pacaud2007}  A $5\sigma$ detection in a map with 12~\ukcmb\ noise per 1\amin\ pixel, the cluster represents the effective cluster mass sensitivity of the instrument.

\section{Conclusions}
\label{sec:conc}

We have presented the design and described the performance of \apex, an instrument primarily designed to observe the SZE in galaxy clusters at 150~GHz.  \apex\ is the result of several years of development and employs robust, scalable technologies in a modular design which have already been adapted and significantly improved for use in other instruments.  The focal plane is a lithographed TES bolometer array with a 280-channel frequency multiplexed SQUID readout system.  The array is cooled by a pulse-tube cooler and helium sorption refrigerator to an operating temperature of 280~mK.

Data from observations were used to analyze detector yield, beams, bandwidth, optical efficiency, optical loading, and sensitivity.  The detector beams show good uniformity across the array and a low level of crosstalk between adjacent channels.  The atmosphere and telescope are the primary optical load.  Despite a high peak optical efficiency, the sensitivity of the type-1 detectors is limited by a narrow bandwidth.  The median sensitivity per type-1 channel is 890~\ukcmbs\ (NE$y$ of 3.5\sci{-4}~$\sqrt{s}$).  In 2009, we implemented a type-2 detector sub-array with an improved cavity design with a \bso\ backshort and a median sensitivity of 530~\ukcmbs\ (NE$y$ of 2.2\sci{-4}~$\sqrt{s}$).

\apex\ has observed for a total of 875~h since 2007.  We have produced high signal to noise images of x-ray selected clusters and produced low noise maps of fields chosen to overlap with external data sets.   The cluster map of Fig.~\ref{fig:bullet} and modeling of the Bullet cluster are presented in detail in Ref.~\onlinecite{halverson2009}.  A power spectrum analysis of our observations of the XMM-LSS\cite{pierre2004} field is given in Ref.~\onlinecite{reichardt2009}. In Ref.~\onlinecite{nord2009}, \apex\ and x-ray observations of Abell 2163 are used to deproject the thermal structure of the intra-cluster medium.  Reference~\onlinecite{basu2010} details a similar analysis of Abell 2204.

Observations were primarily focused on a sample of x-ray selected clusters spanning a wide range of mass and dynamical state.  \apex\ has been used to image 48 clusters. These observations will be used to study the scaling of the SZE signal with masses derived from weak lensing, x-rays, or optical richness for a sample of clusters.

\begin{acknowledgments}
We thank the staff at the APEX telescope site, led by David
Rabanus and previously by Lars-\AA ke Nyman, for their exceptional support, and the machine shop staff of the University of California, Berkeley for their assistance in designing and work in fabrication of the \apex\ receiver system.  We also thank LBNL engineers John Joseph and Chinh Vu for their work on the
readout electronics.  We thank Bryan Steinbach for calculation of AC-biased detector responsivity.

\apex\ is funded by the National Science Foundation under Grant Nos.\ AST-0138348 \& AST-0709497. Work at LBNL is supported by the Director, Office of Science, Office of High Energy and Nuclear Physics (ATL and HS), and by the Director, Office of Science, Office of Basic Energy Sciences, Materials Sciences and Engineering Division (JC, collaboration on development of SQUID multiplexer), of the U.S. Department of Energy under Contract No. DE-AC02-05CH11231. Work at McGill is supported by the Natural Sciences and Engineering Research Council of Canada, the Canadian Institute for Advanced Research, and Canada Research Chairs program.  NWH and MD acknowledge support from Alfred P. Sloan Research Fellowships. CH and DJ acknowledge financial support from the Swedish Research Council.  RK acknowledges partial financial support from MPG Berkeley-Munich fund.  FP acknowledges support from the grant 50 OR 1003 of the Deutsches Zemtrum f\"ur Luft- und Raumfahrt.

\end{acknowledgments}

\bibliography{master_references}

\end{document}